# Predictive and feedback signals differently shape the formation of language representations

Shuguang Yang [1, 2], Shaoyun Yu [3, 4], Xin Jiang [1, 2], Suiping Wang [1,2], Gangyi Feng [3, 4]*

1. School of Psychology, South China Normal University, Guangzhou 510631, China
2. Philosophy and Social Science Laboratory of Reading and Development in Children and Adolescents (South China Normal University), Ministry of Education, Guangzhou 510631, China
3. Department of Linguistics and Modern Languages, The Chinese University of Hong Kong, Hong Kong SAR, China
4. Brain and Mind Institute, The Chinese University of Hong Kong, Hong Kong SAR, China

* Corresponding author

Gangyi Feng
Brain and Mind Institute
Department of Linguistics and Modern Languages
The Chinese University of Hong Kong, Hong Kong SAR, China.
Email: g.feng@cuhk.edu.hk

## Funding

The work described in this paper was supported by grants from the National Natural Science Foundation of China (Project No. 32322090 to Gangyi Feng), the Research Grants Council of the Hong Kong Special Administrative Region, China (Project No.: 14614221, 14612923, 14621424, and C4001-23YF to Gangyi Feng), and Key Research and Development Program of Guangdong, China (grant number: 2023B0303010004 to Suiping Wang).

## Conflict of interest statement

The authors declare no competing financial interests.

**Abstract**
Adults vary greatly in how effectively they learn a new language, but the signals driving the learning processes and individual differences remain unclear. Over seven days, we tracked behavioral learning and collected fMRI data on day 1 from 102 adults as they learned an artificial language with corrective feedback. We trained matched transformer models to predict upcoming linguistic elements, learn grammaticality from corrective supervision or reinforcement-like reward signals in the feedback, and compared their internal representations with human brain activity. Representations derived from the prediction-focused model accounted for the largest share of unique neural variance at the group level, despite the human task being feedback-based. Across model-training stages, alignment between the predictive model and neural representations shifted from sensory to higher-order language and associative networks, indicating a model abstraction process. Conversely, neural patterns related to the feedback model were most useful for predicting individuals' future generalization outcomes on Day 7. Reward signals did not significantly account for learning variance of either type. These findings support a multi-signal model of adult language learning, in which prediction shapes a common neural learning architecture across learners, whereas feedback-related mechanisms better explain individual differences over time.

**Significance Statement**
Learning a new language in adulthood is often challenging overall and varies widely in success, yet the signals that drive these outcomes remain debated. Hypotheses emphasize either predicting upcoming input or learning from corrective feedback, but the two are rarely compared directly. We tracked 102 adults learning an artificial language for seven days, recorded their brain activity in the early stages, and compared it with language models trained with different objectives. The prediction-trained model best captured brain patterns shared across learners. Yet brain signals captured by the feedback-trained model better predicted which individuals would generalize better on day 7. These findings suggest that constructing shared representations and achieving successful future generalization place different computational demands on learning.

## Introduction

Adults often spend years learning a new language, yet progress is often slow, effortful, and varies greatly among individuals. Even with similar instruction and exposure, some learners rapidly acquire linguistic regularities and generalize them to new contexts, whereas others improve more slowly or reach only modest proficiency (Kidd et al., 2018; Morgan-Short, 2020). Understanding the language learning process and its variability is key to understanding human learning, as it affects not only language acquisition but also how the adult brain builds new, structured knowledge from experience.

A longstanding debate centers on whether adult language learning is driven primarily by a general learning process or by multiple processes or signals that contribute differently depending on the learning demand (Minda et al., 2024; Morgan-Short, 2020; Hamrick et al., 2018; Ullman, 2004; Ashby et al., 1998). We focus on two interacting sources of learning signals, distributional regularities and explicit correctness feedback. Both can support key putative learning processes such as prediction, error-driven updating, and generalization (Bovolenta & Marsden, 2022). One highlights prediction-based learning, where learners track distributional structure in the input and form expectations about upcoming sounds, words, and syntactic structures. This enables them to build representations that support generalization across different contexts and time frames (Caucheteux et al., 2023; Heilbron et al., 2022; Pelucchi et al., 2009; Saffran, 2002; Saffran et al., 1996). The other emphasizes learning from explicit correctness feedback, which can inform learners about the appropriate response and help refine language representations (Opitz et al., 2011). Both perspectives are plausible and may interact or overlap, but it remains unclear how these learning signals shape the emergence of language representations and whether they matter equally for the common neural organization versus individual differences among learners. Learning involves shared structures and individual variability in how learners extract and use environmental information. Despite encountering the same linguistic signals, learners differ in their sensitivity to predicting regularity and in their use of feedback, affecting outcomes. We aimed to examine whether prediction- and feedback-driven learning pressures contribute equally to shared neural representations and individual differences in future generalization outcomes.

This uncertainty persists in part because the two learning signals are usually studied separately. Prediction-based learning is often examined using statistical learning or self-supervised language modeling (Bovolenta & Marsden, 2022; Ito et al., 2018; Rabagliati et al., 2015), whereas feedback-based learning is often examined through tasks providing explicit information about response correctness (Arbel et al., 2014, 2017). These different approaches make it difficult to compare the representational consequences of distinct learning objectives within a common learning environment. Here, we used matched model architectures and identical artificial-language materials to examine whether different optimization objectives produce distinguishable language representations, whether different stages of model optimization correspond differently to components of the emerging neural geometry, and whether neural information captured by these model representations differentially predicts future generalization success across individuals.

We addressed these questions using a unified approach to compare candidate learning signals in a common computational and representational space. Transformer models are particularly useful for this purpose. Classical symbolic, feature-engineered and statistical models have provided important mechanistic insights, but they typically rely on hand-crafted representations or modular pipelines that are less suited to capturing end-to-end representational change during learning (Elman, 1990; Naselaris et al., 2011; Riesenhuber & Poggio, 1999). By contrast, transformer models can be trained with different objectives on the same corpus, yielding directly comparable internal representations within a shared architecture. Recent work has also shown that such representations align with human neural activity across the cortical hierarchy during language processing (Caucheteux et al., 2022, 2023; Goldstein et al., 2022). This makes them a powerful tool for testing where, when, and to what extent distinct learning objectives resemble the representational geometry of the human brain.

A key conceptual point is that the current models focus on learning signals rather than fully implementing entire theoretical frameworks. Specifically, we operationalize the prediction objective as next-token prediction and the feedback objective as correctness-supervised grammaticality judgment using a trial-wise feedback structure. Both models make predictions and learn from errors, but differ in their prediction targets and supervision. Their comparison tests the representational consequences of next-token prediction and correctness-supervised classification. The prediction objective imposes internal pressure to anticipate upcoming structure from distributional regularities, whereas feedback imposes external pressure to update responses based on success and error signals. Therefore, they should not be viewed as direct, one-to-one implementations of predictive coding theory (Friston, 2005) or algorithmic reinforcement learning (Sutton & Barto, 2018) in their strongest forms. Instead, they serve as controlled approaches to test whether representations optimized for prediction and feedback objectives relate differently to human neural activity and behavior. This distinction matters because prediction can either drive language learning or emerge as a byproduct of it (Chang et al., 2006, 2013). Thus, this comparative framework allows us to manipulate learning objectives while keeping the model architecture and training materials constant, enabling us to measure how the resulting internal representations correspond to neural activity during human learning. It also lets us test whether mechanism-related neural information better predicts individual learning outcomes.

We applied this framework to a seven-day artificial-language learning study using Brocanto2, a miniature artificial language with a controlled lexicon and grammar embedded in a meaningful board-game context (Morgan-Short et al., 2010, 2012). We tracked behavioral learning in 102 participants over the full training period and collected fMRI data on Day 1, while participants learned Brocanto2 with corrective feedback. In parallel, we trained three GPT-style models with the same architecture but different objectives (Fig. 1). One was a prediction model (GPT-P) optimized for next-token prediction, another was a feedback model (GPT-F) optimized for grammaticality judgment with corrective feedback, and the third was a combined model (GPT-PF) optimized for both tasks. We also trained a model optimized for reward signals (GPT-R) as a control. With architecture and linguistic input held constant, these models serve as controlled computational probes of which learning objectives yield representations that better account for human neural patterns. Such correspondence can constrain computational accounts of learning without assuming that humans share the models' architecture or learning algorithms. We then aligned trained model representations with fMRI data using representational similarity analysis (RSA) and variance partitioning, tracked model-stage-specific representational updates against the fixed Day-1 neural reference, and built individualized learner models that combined early neural and behavioral information to predict Day-7 generalization outcomes.

Previous studies have shown that predictive language models capture aspects of human neural representations during language processing, but whether different learning objectives account for neural representations during language learning remains unclear (Caucheteux & King, 2022; Goldstein et al., 2022; Schrimpf et al., 2021). We therefore compared the neural correspondence of models optimized for next-token prediction and correctness-supervised classification during early language learning. Given the hierarchical organization of linguistic representations across sensory, language and association networks, we further examined how model-brain correspondence varied across neural systems and model-training stages (Caucheteux et al., 2023; De Heer et al., 2017; Mischler et al., 2024), using Day-1 neural representations as a fixed reference. Finally, building on evidence that early neural activity predicts later language-learning outcomes, we investigated whether neural information captured by the different models could explain individual differences in subsequent generalization (Feng et al., 2021; Song et al., 2026). Together, these analyses address a broader question of whether the computational objectives that best capture shared neural representations also provide the most informative account of individual learning success.

By combining prediction- and feedback-based objectives within a common computational and neuroimaging framework, this study reframes the debate over adult language learning from a competition between single mechanisms to a question of functional dissociation. We demonstrated that prediction-based representations best explain the early-learning neural geometry shared across learners, whereas feedback-related neural information best predicts individual differences in future generalization. These findings suggest potentially complementary explanatory roles for the information captured by the two objectives. More broadly, our approach establishes model-brain alignment as a computational probe for testing how adults acquire new structured knowledge and why learners exposed to the same language input arrive at different outcomes.

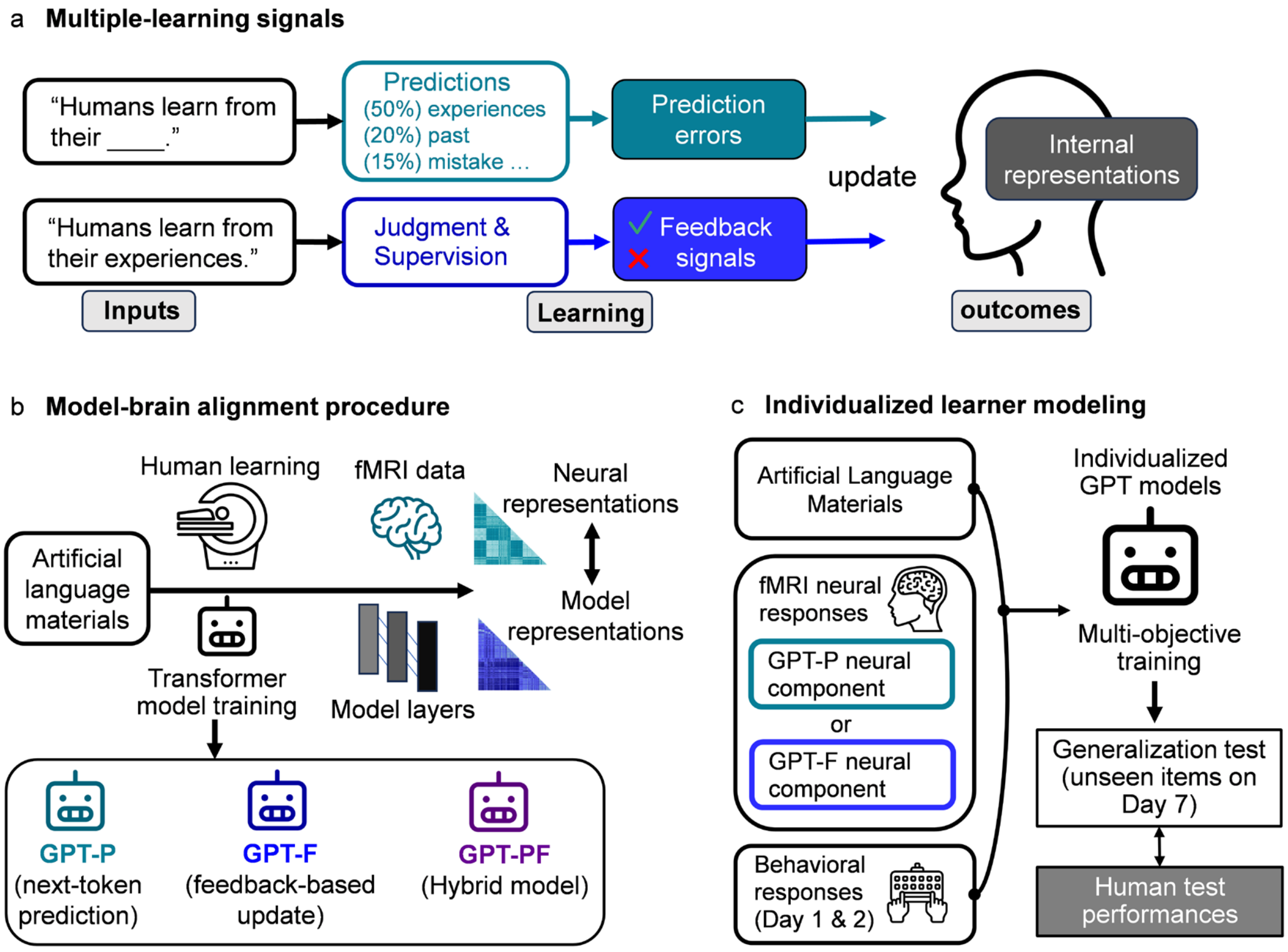

**Fig. 1. Overview of the study framework and procedure. a**, Hypothesized learning mechanisms supporting human language learning. The framework compares two candidate signals: prediction, based on next-element prediction among alternatives, and feedback-based learning, based on outcome signals (correct or incorrect). Both update the internal representations of language. **b**, Model-brain alignment procedure. Participants learned an artificial language over seven days of training while fMRI data were collected on Day 1, and parallel GPT models were trained on the language materials. Three models were compared: GPT-P (next-token prediction), GPT-F (feedback-based learning), and GPT-PF (both). RSA compared model layers' representations with neural responses. **c**, Individualized learner modeling. To explain individual differences in learning, participant-specific GPT models combined artificial-language materials with early behavioral responses and model-captured neural information. These models predicted Day-7 generalization performance on unseen items. This analysis tested whether neural information captured by GPT-P or GPT-F representations differed in its prospective relevance for later individual learning outcomes.

## Results

### Human and model language-learning performance

We collected fMRI data on Day 1 from 102 adults performing an artificial language-learning task that required grammaticality judgments with trial-by-trial corrective feedback. Participants then completed six additional days of behavioral training using the same paradigm, followed by a Day 7 generalization test that assessed grammaticality knowledge on unseen items (Fig. 2a).

To test the candidate learning mechanisms, we trained three small-scale GPT-style models with identical architectures and hyperparameters, differing in their training objectives and corresponding supervision. Each model used a two-layer decoder-only architecture, comprising one token-embedding layer and two hidden layers. GPT-P was trained for next-token prediction using cross-entropy loss. GPT-F was optimized for grammaticality classification using supervised cross-entropy loss, with grammaticality labels providing the corrective information available during human training. GPT-PF combined both objectives by jointly optimizing prediction and feedback objectives with dynamic loss weighting (see Methods). By holding the model backbone constant while varying the learning objectives and their associated supervision, this framework allowed us to compare the representational consequences of next-token prediction, correctness-supervised classification, and their combination. Because these objectives necessarily involved differences in training inputs and supervisory targets, the comparisons do not isolate the learning objectives from all other aspects of training experience (e.g., no feedback in GPT-P).

All three models acquired their respective tasks within 200 training epochs, although validation performance followed non-monotonic trajectories (Fig. 2c). At their best-performing checkpoints, GPT-P achieved a generation success rate of 91%, whereas GPT-F reached 100% accuracy on the grammaticality-judgment test (GJT). The combined GPT-PF model achieved a higher peak generation success rate (96%) than GPT-P while maintaining high GJT accuracy (97%). These results indicate that both optimization objectives supported successful acquisition of their respective tasks, and that combining the objectives was compatible with high performance on both. In addition, grammaticality judgment in both GPT-F and GPT-PF exceeded human learners' Day-7 peak accuracy (79%; Fig. 2b and Supplementary Fig. S1).

To examine how the models' capabilities emerged, we tested whether changes in their internal layer representations predict changes in their task performance across training checkpoints (see Methods and Fig. S2). We computed representation-update RDMs and performance-update RDMs, as well as their associations, for all three models (Fig. 2d). In GPT-P, we found significant associations across all layers, though weaker than in GPT-F. In GPT-F, the associations were robustly positive and strongest in the two hidden layers, indicating tighter coupling between representation updating and performance improvement. In GPT-PF, representational updates were positively related to both sentence generation and grammaticality judgment performance, with the effects increasing over layers. Thus, learning in all three models involved systematic reorganization of internal representations relevant to the tasks.

We next compared representational geometry across models (Fig. 2e). Among model pairs, GPT-P and GPT-F showed the lowest similarity between corresponding hidden layers ($r_{\text{layer1}} = 0.081$, $r_{\text{layer2}} = 0.076$), indicating that the different training objectives were associated with substantially different representational geometries. Although GPT-PF shared the same architecture as the two single-objective models, its layer representations were more similar to GPT-F ($r_{\text{layer1}} = 0.357$, $r_{\text{layer2}} = 0.543$) than to GPT-P ($r_{\text{layer1}} = 0.298$, $r_{\text{layer2}} = 0.272$), especially in Layer 2.

We then examined how GPT-P and GPT-F aligned with human learning behavior (i.e., responses) across model training (Fig. 2f). For this analysis, we constructed a mean behavioral-response RDM from human judgment patterns across the seven training days and compared it with model RDMs extracted from four checkpoints sampled at comparable intervals of performance improvement along each model's training trajectory. These checkpoints divided each model's training trajectory into comparable stages of improvement. In Layer 1, GPT-P showed stronger model-behavior alignment at the first two checkpoints ($t_{(101)} = 3.06$, $p = 2.82 \times 10^{-3}$ and $t_{(101)} = 4.26$, $p = 4.56 \times 10^{-5}$), did not differ reliably from GPT-F at the third checkpoint ($t_{(101)} = 0.95$, $p = 0.34$), and was weaker than GPT-F at the final checkpoint ($t_{(101)} = -12.13$, $p = 1.99 \times 10^{-21}$). In Layer 2, GPT-P remained more strongly aligned through the first three checkpoints ($t_{(101)} = 4.98$, $p = 2.60 \times 10^{-6}$, $t_{(101)} = 5.07$, $p = 1.80 \times 10^{-6}$, and $t_{(101)} = 2.77$, $p = 6.60 \times 10^{-3}$), whereas GPT-F showed stronger alignment at the final checkpoint ($t_{(101)} = -9.30$, $p = 3.11 \times 10^{-15}$). These results demonstrate that model-behavior correspondence varies across stages of model training. Relative to the average human behavioral reference, early-to-intermediate GPT-P representations showed stronger alignment, whereas later GPT-F representations showed stronger alignment as model performance approached its maximum. Thus, the two objectives captured different representational information at different stages of model training.

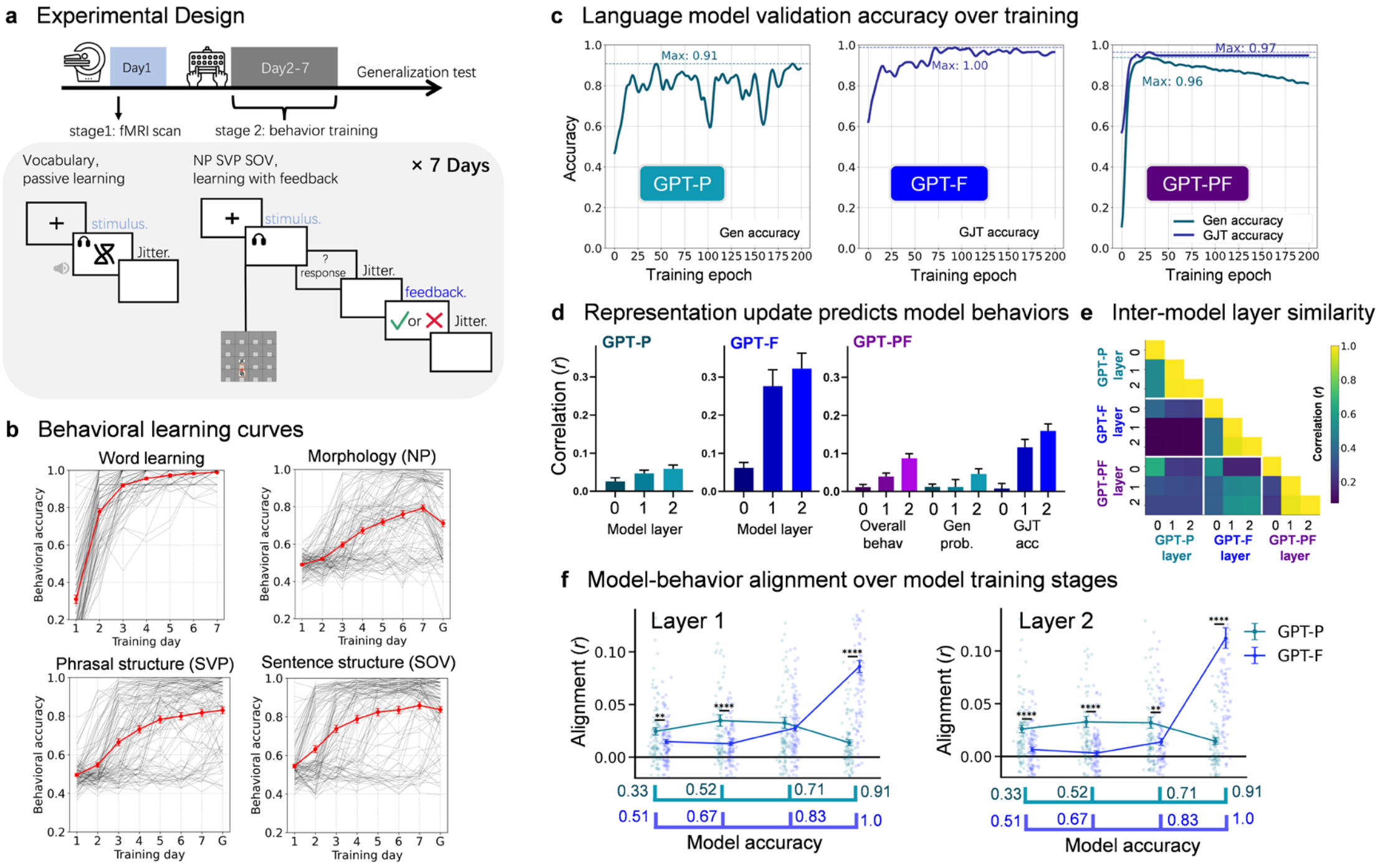

**Fig. 2.** Experimental design, behavioral learning curves, and model-behavior alignment. **a**, Seven-day training and neuroimaging procedure. Participants underwent fMRI on Day 1 during vocabulary and grammar training, followed by behavioral training on Days 2-7, ending with a generalization test. **b**, Behavioral learning curves over the 7 training days for word, morphology, phrasal, and sentence structure. Gray lines represent individual participants; red lines show the group means. G = generalization scores. **c**, Validation performance of the three models across training epochs. For GPT-P and GPT-PF, performance is displayed as the rate of generating novel grammatical sentences ("Gen accuracy"); for GPT-F and GPT-PF, grammaticality judgment test (GJT) accuracy is shown. **d**, Associations between model representational and behavioral change are shown with bar plots of Spearman correlations between representational-update RDMs and performance-update RDMs across layers and models. **e**, Similarity of internal representations across models. The heat map displays layer-wise correlations between model RDMs. **f**, Model-to-human-behavior representation alignment. Mean RSA alignment (*r*) between model representations and human behavioral patterns is shown across four model checkpoints. Dots indicate individual participants, and error bars indicate SEM. Asterisks indicate paired *t*-test results for the comparison between GPT-P and GPT-F at each stage. **, $p < 0.01$; ****, $p < 0.0001$.

## Prediction-optimized representations show the strongest group-level neural correspondence

We next examined how closely the representational geometry of the three models aligned with human neural RDMs during learning. To this end, we used whole-brain searchlight RSA (Kriegeskorte, 2008; Stolier & Freeman, 2016) with multiple regression (Fig. 3a) to reveal the unique contribution of each model. Layer-wise RDMs from the models were fitted to predict fMRI-derived RDMs while controlling for three nuisance RDMs (i.e., an untrained GPT model, stimulus length, and block identity). We further complemented whole-brain RSA with variance partitioning (Groen et al., 2018) and network-wise RSA across 12 canonical functional brain networks (Ji et al., 2019).

All three models showed reliable alignment with human neural representations (Fig. 3b; see Fig. S4 for layer-wise RSA maps). However, the extent and distribution of this alignment differed across models. Although the human task focused on grammaticality judgments with feedback, the GPT-P model accounted for the greatest amount of neural variance (Fig. 3c). To test if GPT-P's neural correspondence relied on its training set, we trained GPT-P(L) on 288 grammatical sequences, matching the human task size but with grammatical-only input. GPT-P(L) showed the same model-brain alignment as GPT-P (Supplementary Fig. S5). This alignment also extended beyond cortical regions to subcortical structures, including the caudate, nucleus accumbens, and globus pallidus. By contrast, the feedback model (GPT-F) aligned with human data closely in canonical perisylvian language regions (Hagoort, 2019) and the default mode network (Menon, 2023), whereas the integrated model (GPT-PF) showed additional alignment only in early sensorimotor cortices.

These between-model differences were further supported by variance partitioning. GPT-P accounted for the largest share of unique neural variance across model layers (Fig. 3c). We then examined whether the models differed in their alignment with neural representations extracted from 12 canonical brain networks. Consistent with the overall variance-partitioning results, GPT-P showed significantly higher alignment than the other models in most networks (8 of 12; wilcoxon-tests; Fig. 3d). GPT-F did not significantly outperform GPT-P in most networks. These converging results indicate that prediction-based representations provide the strongest group-level account of neural patterns during human language learning.

To assess robustness to random initialization, we independently trained 10 instances each of GPT-P and GPT-F and repeated the representational and whole-brain RSA analyses (Supplementary Fig. S3). Internal representational geometries were reproducible across runs, with mean inter-seed RDM correlations of 0.710 and 0.706 for GPT-P Layers 1 and 2, respectively, and 0.762 and 0.775 for GPT-F. Neural-map reproducibility, however, differed substantially between objectives. GPT-P showed high pairwise spatial correlations between unthresholded whole-brain RSA maps in both layers (Layer 1: mean $r = 0.868$, 95% CI [0.793, 0.943]; Layer 2: mean $r = 0.859$, 95% CI [0.776, 0.942]). Correlations between each run's map and the mean map of the other nine runs were similarly high (Layer 1: mean $r = 0.925$; Layer 2: mean $r = 0.919$). Seed-averaged GPT-P RDMs also reproduced broad neural alignment in both layers at FWE-corrected $p < 0.05$.

Despite reproducible internal representations, GPT-F showed lower spatial reproducibility of neural alignment (Layer 1: mean $r = 0.113$, 95% CI [0.003, 0.224]; Layer 2: mean $r = 0.130$, 95% CI [0.013, 0.247]). Seed-averaged GPT-F RDMs yielded significant alignment in Layer 1 but not Layer 2 at the same corrected threshold. Within-GPT-P map similarity also exceeded GPT-P–GPT-F similarity in both layers (Layer 1: $\Delta r = 0.979$; Layer 2: $\Delta r = 1.013$; both permutation $p = 0.001$). These results establish the reproducibility of the principal GPT-P alignment across independent initializations while indicating greater variability in GPT-F neural correspondence.

To examine how different computational uses of outcome information relate to neural representations, we trained GPT-R, a policy-gradient reinforcement-learning counterpart to the supervised GPT-F classifier, and GPT-PR, which combined next-token prediction with policy-gradient learning. These models allowed us to examine whether successful grammaticality performance was sufficient to recover human representational geometry. We conducted the same multiple regression analysis, replacing GPT-F and GPT-PF's layer-wise RDMs with those of GPT-R and GPT-PR, while holding the other regressors constant. Although GPT-R achieved similar task accuracy to GPT-F, its representations did not significantly align with human neural activity (Fig. 3b, dashed box; Supplementary Fig. S6). This finding persisted across 10 independently initialized GPT-R instances: peak task accuracy remained high (mean = 96.04%, SD = 4.15%), and internal RDMs were reproducible (Layer 1: mean inter-seed $r = 0.733$; Layer 2: $r = 0.695$), yet seed-averaged RDMs yielded no significant neural alignment in either layer (Supplementary Fig. S3). Thus, the absence of significant GPT-R alignment persisted after averaging across independent initializations despite successful task learning.

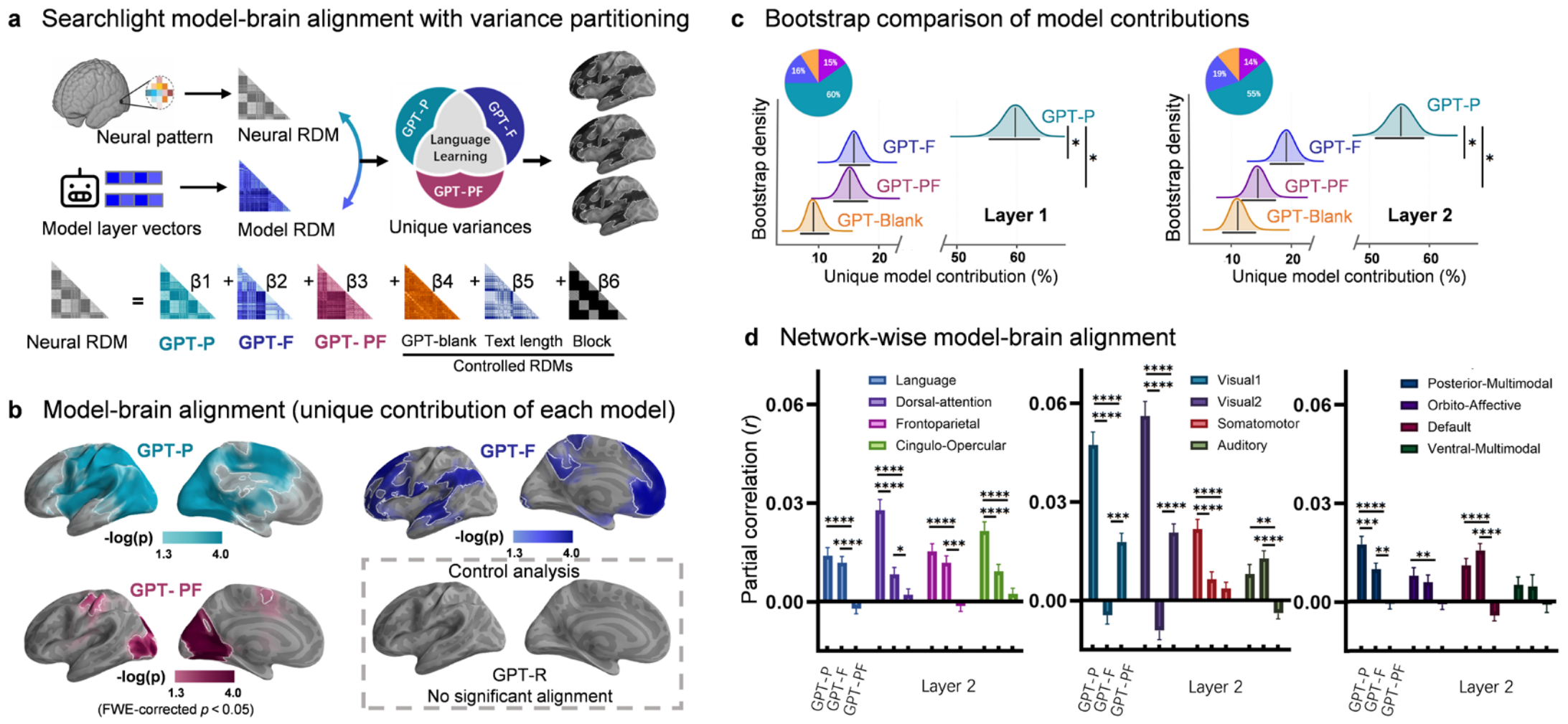

**Fig. 3. Whole-brain and network-wise RSA dissociate the contributions of prediction and feedback-based learning. a,** Schematic overview of the whole-brain searchlight RSA and variance-partitioning procedure. At each searchlight location, neural RDMs were regressed against model RDMs to estimate the contributions of the models to neural representations, controlling for an untrained GPT baseline (GPT-Blank), text length, and experimental block. **b,** Whole-brain searchlight RSA results for models' layer 2. In the control analysis (dashed box), the policy-gradient reinforcement model (i.e., GPT-R) showed no significant alignment, despite robust model performance (see Fig. S6). Brain maps were thresholded using TFCE with 10,000

permutations and family-wise-error corrected $p < 0.05$. **c**, Bootstrap distributions of the relative unique contribution of each model to neural representational variance. Bootstrap distributions of explained variance for each model RDM across first and second layers. Each bootstrap sample's $R^2$ for each model RDM is shown as a percentage of the total $R^2$ from GPT-P, GPT-F, GPT-PF, and GPT-Blank. Density plots show resampled contribution estimates, with horizontal lines for central tendency and asterisks indicating significant differences. **d,** Network-wise RSA across 12 canonical functional networks for the models' layer 2. Bars show partial correlations between neural and model RDMs for each network, controlling for the other RDMs. Prediction-based representations showed the strongest alignment across sensorimotor, language, and association networks, whereas feedback-based representations contributed more selectively. *, $p < 0.05$; **, $p < 0.01$; ***, $p < 0.001$; ****, $p < 0.0001$.

Together, these findings indicate that models optimized for next-token prediction capture a broader and more reproducible component of the neural representational geometry measured during early language learning than models optimized for correctness-supervised classification or policy-gradient learning. The policy-gradient control further shows that successful grammaticality classification alone does not guarantee model-brain correspondence. These results support a role for predictive optimization in accounting for neural representations during language acquisition, without establishing that prediction is necessary or sufficient for human learning.

**Model-brain correspondence shifts across model training from sensory to higher-order neural systems**

To examine how model-brain correspondence changes across model training, we performed RSA on representational updates rather than on cumulative checkpoint representations alone. Because human fMRI data were available only from Day 1, the neural RDMs served as a fixed reference for representational geometry during early human learning. We therefore compared successive model-training stages with the same Day-1 neural reference to determine whether different stages of model optimization corresponded to different components of this neural geometry. Here, "early," "middle," and "late" denote successive model-training stages defined by checkpoints; the same participant-specific Day-1 neural RDMs were used as the reference for every stage.

We sampled four checkpoints at similar intervals of improvement in validation accuracy, dividing training into three stages (Fig. 4a; see Methods for details). For each stage, we computed representational change vectors and derived RDMs summarizing the geometry of these updates. Checkpoint RDMs capture the cumulative representational state at each training stage; update RDMs characterize changes in hidden-state geometry between successive checkpoints. This allows us to examine which components of the fixed neural reference align more closely with model representational changes at different stages.

We found that the neural components corresponding to model representational updates differed across model-training stages. The pattern shifted from lower-level sensory systems at earlier model stages toward higher-order language and associative systems at later stages (Fig. 4b). For GPT-P, alignment gradually shifted from primary sensorimotor regions to the language and default mode networks across training (Fig. 4b, lower panel), indicating that the model might first capture relatively low-level or surface representational structure and only later acquire representational structures resembling those in higher-order systems.

To quantify representational alignment trajectories, we assessed changes in alignment across 12 brain networks. In GPT-P (Fig. 4c), alignment decreased across training stages in lower-order regions such as the secondary visual network, while increasing in higher-order systems such as the language and associative networks (Fig. 4d-e). GPT-F showed a qualitatively similar but weaker and delayed pattern. Alignment first appeared in lower-order systems at the middle model-training stage, with significant effects in the secondary visual network and the adjacent ventral multimodal network. Subsequently, this alignment extended toward higher-order cortical systems at the late model-training stage.

To quantify differences across model-training stages, we fitted a linear regression to each participant's partial alignment values across the three stages, separately for each model and network. These slopes describe changes in alignment across ordered model-training stages against a fixed Day-1 neural reference. GPT-P exhibited significantly steeper positive slopes than GPT-F in the language and default mode networks (Wilcoxon tests: $p_{\text{Language}} = 2.44 \times 10^{-7}$, $p_{\text{DMN}} = 1.64 \times 10^{-11}$; Fig. 4e), indicating a larger increase in correspondence with these networks across the sampled model stages.

In summary, representational updates from successive training stages showed different patterns of correspondence with the fixed Day-1 neural reference. For GPT-P, this correspondence shifted from sensory networks toward language and association networks, with a stronger increase in the latter systems than was observed for GPT-F. These findings suggest that the two optimization objectives shape the emergence of representational structure differently, with later model updates capturing aspects of the higher-order neural geometry already present during early human language learning.

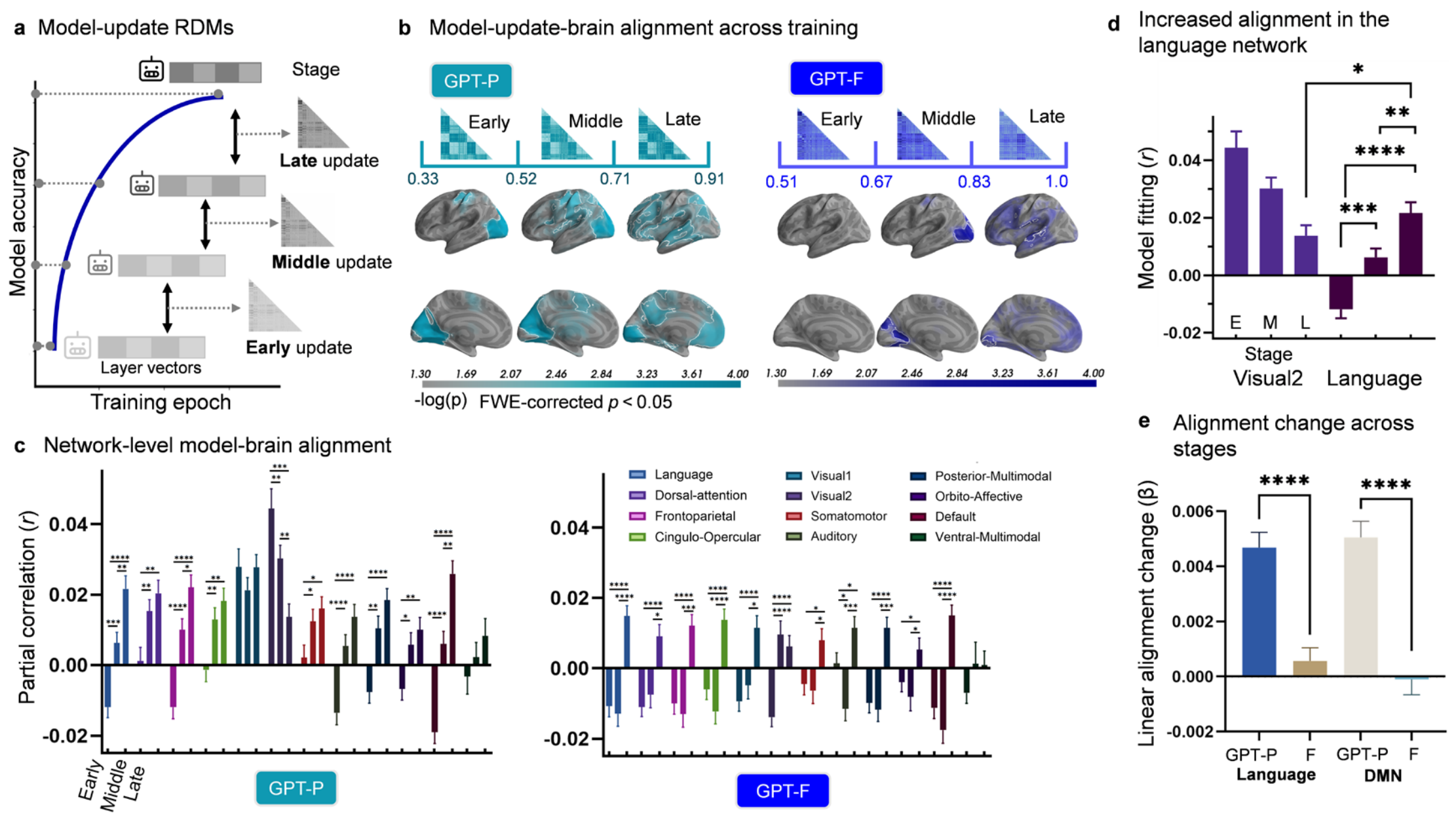

**Fig. 4. Model-brain representational alignment shifts from sensorimotor to language and higher-order systems. a,** Schematic showing representation-update-based brain-model alignment analysis. Vector representations were sampled at multiple points during model training, and changes were measured as the differences between adjacent checkpoints to construct the update RDM. This captures how internal model geometry updates across training stages. Early, middle, and late stages refer to model training. All stage-specific model RDMs were compared with the same neural RDMs derived from the Day-1 fMRI session. **b,** Whole-brain RSA results for representational updates in the first layer of GPT-P and GPT-F across three successive training stages. The numbers below the brain maps show model performance for sampled checkpoints. Brain maps are thresholded using TFCE with 10,000 permutations and FWE-corrected $p < 0.05$. **c,** Network-wise RSA of representational updates across training stages for both models. Bars show partial correlations between neural RDMs and stage-specific update RDMs across the 12 functional networks. **d,** Direct comparison of GPT-P update alignment in the secondary visual (Visual2) and language networks. **e,** Linear slopes fitted to each participant's stage-wise update alignment values. GPT-P showed significantly larger increases than GPT-F in both the language and default mode networks, indicating a stronger late-emerging shift towards higher-order representational systems. *, $p < 0.05$; **, $p < 0.01$; ***, $p < 0.001$; ****, $p < 0.0001$.

## Individualized neurocomputational models identify the predictors of individual differences

We further examined which learning-related signals better explain individual differences in final learning outcomes. To address this question, we developed individualized GPT models trained on each participant's early behavioral and neural responses (Fig. 5a). Each model was trained with three sources of information: next-token predictions (for learning the language materials), individual learners' grammaticality judgment responses (for learning the human task), and individual neural features captured by the models. The individualized models shared the same architecture, training objectives, loss weights, and optimization procedure, differing only in the neural information incorporated during training. This comparison tested the predictive value of neural information associated with GPT-P and GPT-F representations, rather than differences between learning mechanisms implemented by the individualized models.

The individualized models reproduced participants' early behavioral responses, while incorporating neural objectives increased correspondence with the neural features used to constrain model training (Fig. 5b-c). We next examined whether these models generalized prospectively to unseen Day-7 items, which were excluded from fitting and checkpoint selection. Early behavioral information alone provided modest predictive information about Day-7 generalization ($r = 0.27$). Incorporating GPT-F-aligned neural information yielded a stronger association between predicted and observed generalization scores ($r = 0.43$, $p = 6.86\times10^{-6}$; Fig. 5d). A direct bootstrap comparison confirmed that this model outperformed the behavioral-only model ($p = 0.0009$; Fig. 5e). By contrast, incorporating GPT-P-aligned neural information did not significantly improve prediction over behavioral information alone ($p = 0.17$). The GPT-F-aligned model also showed higher predictive performance than the GPT-P-aligned model ($p = 0.047$), indicating that the two neural information sources differed in their prospective utility within the individualized modeling framework.

This advantage was not attributable to differences in voxel count between the brain masks. The GPT-F mask initially contained fewer voxels than the GPT-P mask. To rule out a simple volumetric explanation, we conducted a control analysis in which the statistical threshold for the feedback-region mask was adjusted to yield a voxel count closely matched to that of the prediction-region mask (Supplementary Fig. S7d). Using this voxel-matched mask, we repeated the full individualized modeling procedure. The GPT-F-aligned model continued to significantly predict Day 7 generalization performance ($r = 0.35$, $p = 3.34 \times 10^{-4}$; Supplementary Fig. S7d).

To distinguish model-captured neural information from anatomical region selection, we used cross-validated PLS regression to estimate neural components predicted by GPT-P and GPT-F representations (Supplementary Fig. S7a). After cross-validated adjustment for pre-training nonverbal intelligence (TONI-4), GPT-F-informed predictions remained associated with Day-7 generalization using the original masks (Spearman $r = 0.34$, $p = 4.57 \times 10^{-4}$) and a common language-network mask for both models ($r = 0.36$, $p = 1.90 \times 10^{-4}$). Models using multiple-demand or visual-network responses predicted adjusted generalization ($r = 0.29$, $p = 0.0036$; $r = 0.23$, $p = 0.0184$, respectively), but neither significantly improved prediction over the behavioral-only model (Supplementary Fig. S7c-g).

We additionally evaluated simpler predictive baselines using early neural and behavioral measurements (Supplementary Fig. S9). Nested cross-validated ridge regression significantly predicted Day-7 generalization using GPT-P-aligned features (Spearman $r = 0.26$, $p = 0.008$) and voxel-matched GPT-F-aligned features ($r = 0.27$, $p$

= 0.006). SVR predictions did not reach significance (GPT-P: $r = 0.10$, $p = 0.338$; voxel-matched GPT-F: $r = 0.18$, $p = 0.075$). These results demonstrate that early behavioral and neural information also supports prospective prediction using conventional regularized regression. Because the ridge and SVR analyses evaluated held-out participants, whereas the individualized models were evaluated on unseen items within participants whose early data had been used for model fitting, the magnitudes of their correlations cannot be interpreted as a direct comparison of predictive architectures. The two analyses instead provide complementary evidence for the predictive information contained in early learning measurements.

A complementary analysis using neural constraints associated with late-stage model-update representations likewise showed that GPT-F-related constraints improved generalization prediction over both behavioral alignment ($p = 0.0267$) alone and GPT-P-related constraints ($p = 0.0497$; Supplementary Fig. S8). Here, “late stage” refers to model training; all neural data were collected on Day 1.

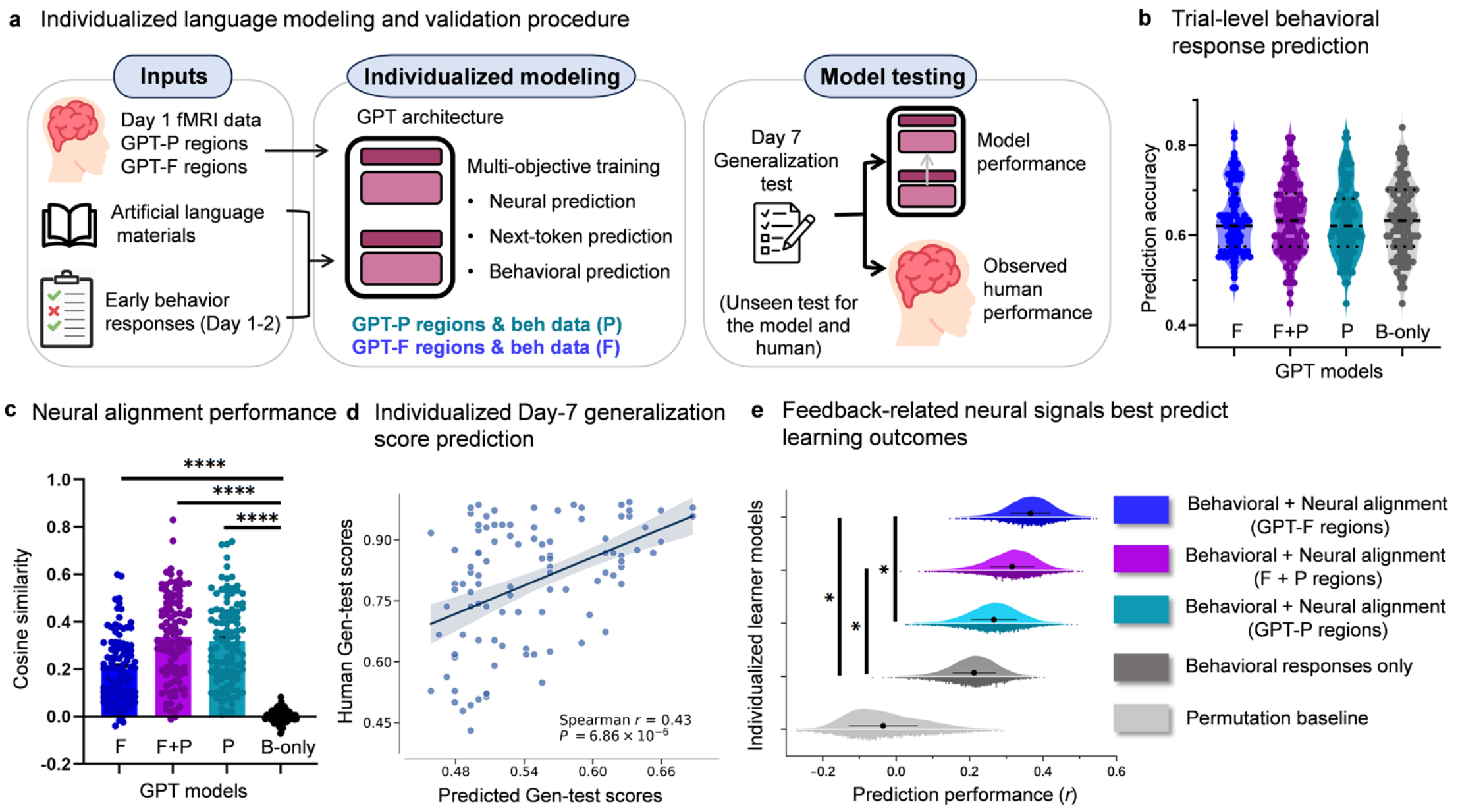

**Fig. 5. Individualized models predict later generalization from early behavioral and neural data. a,** Schematic of the individualized modeling framework. Participant-specific models were trained on next-token prediction (using Brocanto2 materials), early behavioral responses, and Day 1 fMRI-derived model-aligned neural features. **b,** Trial-level behavioral response prediction across individualized models. **c,** Cosine similarity between model representations and participant-specific neural representations. Models trained with neural objectives showed greater neural alignment than the behavioral-only model. **d,** Prediction of Day 7 generalization scores by the best-performing individualized model. **e,** Bootstrap distributions of generalization-prediction performance. The model incorporating GPT-F-aligned neural information outperformed both the behavioral-only model and the model incorporating GPT-P-aligned neural information. The latter did not significantly outperform the behavioral-only model. *, $p < 0.05$; ****, $p < 0.0001$.

Together, these analyses indicate that early neural information captured by feedback-supervised representations provides prospective information about individual differences in later generalization. Its predictive association persists after adjustment for nonverbal intelligence and when neural components are extracted within a common anatomical space. The stronger performance of GPT-F-aligned information within individualized models suggests that the neural characteristics most informative about later learning outcomes need not be those showing the broadest group-level model-brain correspondence.

**Discussion**

Using a seven-day artificial-language-learning paradigm, fMRI recordings during early learning, and the development of group-level and individualized transformer models, we found that prediction and feedback explain complementary aspects of adult language learning. Prediction-based representations best captured neural structure shared across learners, whereas feedback-related neural signals better explained individual differences in later generalization. At the group level, representations optimized for a prediction objective accounted for the largest portion of unique explainable neural variance, even though participants performed a feedback-based learning task. Across model training, both objectives showed a shift in alignment with the fixed Day-1 neural reference from sensory toward higher-order language and associative networks, with a stronger shift for GPT-P. At the individual level, early feedback-related neural responses improved prediction of Day 7 generalization task performance beyond the prediction model and the behavior-only model. Overall, these findings support a multi-signal account of adult language learning. Prediction- and feedback-optimized representations capture different aspects of adult language learning. The former shows stronger correspondence with neural representational structure shared across learners, whereas the latter provides more predictive information about individual differences in subsequent generalization.

The group-level findings are most informative when interpreted in light of the learning objectives implemented. Next-token prediction operationalized a prediction-based learning signal, whereas a grammaticality judgment task with trial-by-trial corrective feedback operationalized a feedback-based learning signal. These objectives should not be viewed as complete implementations of predictive-coding theory or reinforcement learning in an algorithmic sense. However, within this operationalization, our results are consistent with prediction-related computations providing a representational scaffold during even early learning. Prediction has been connected to both language learning and processing, with evidence that learners use distributional regularities to anticipate upcoming linguistic structures and that hierarchical predictive signals can be detected across the cortical language system (Caucheteux et al., 2023; Heilbron et al., 2022; Pelucchi et al., 2009; Saffran, 2002; Saffran et al., 1996). At the same time, prediction has also been argued to arise as a byproduct rather than a driver of learning (Chang et al., 2006, 2013). Our results do not fully settle this broader debate, but they do provide an important constraint. Even in a task designed around explicit correctness feedback, representations optimized for prediction showed the strongest model-brain alignment and explained the largest proportion of unique explainable variance. These findings suggest that optimization for next-token prediction provides a useful computational account of the neural representational structure observed during early human language learning, even when explicit corrective feedback is provided.

This interpretation is also consistent with recent work showing that predictive structure is a major source of alignment between deep language models and human brain activity during language processing (Caucheteux et al., 2022, 2023; Goldstein et al., 2022). In that sense, our findings extend the relevance of prediction beyond mature language comprehension to the formation of new linguistic knowledge. Importantly, the current findings do not imply that prediction alone suffices for effective language learning. Instead, they indicate that representations acquired through predictive optimization capture substantial aspects of the neural geometry observed during early language learning.

Analyses across independent initializations strengthen evidence for prediction-related neural correspondence while clarifying differences between models. GPT-P consistently reproduced the spatial pattern of alignment across runs, whereas GPT-F showed more variable neural correspondence despite reproducible internal representations. The spatial extent and layer specificity of GPT-F alignment therefore require more cautious interpretation. Together with the absence of significant seed-averaged GPT-R alignment, these findings distinguish successful task learning and reproducible model representations from reliable correspondence with human neural geometry.

The GPT-F and GPT-R results distinguish successful task performance from correspondence with human representational geometry. GPT-R achieved high accuracy but showed no significant seed-averaged neural alignment, whereas GPT-F retained alignment in Layer 1. One possible explanation is that explicit category supervision encourages representational distinctions relevant to human learning that the tested reward-optimization procedure does not consistently recover. This remains a hypothesis. Establishing biological plausibility requires a separate account of how learning signals produce synaptic updates; neither a policy-gradient objective nor representational alignment establishes that correspondence (Lillicrap et al., 2020). The findings therefore constrain these particular models without excluding reinforcement-learning contributions to language learning. Similarly, the predictive utility of GPT-F-associated neural information should be interpreted at the representational level, without identifying a specific biological learning rule (Kriegeskorte & Kievit, 2013).

The stage-specific analyses characterize how model representational updates increasingly correspond to different components of the early-learning neural reference. Updates during earlier model-training stages aligned primarily with sensory-motor regions, consistent with sensitivity to surface stimulus properties (Brincat et al., 2018). Updates during later model-training stages showed greater correspondence with language and association networks, potentially reflecting increasingly abstract model representations (Courellis et al., 2024; Johnston & Fusi, 2023). This interpretation fits prior evidence that language comprehension is organized along a cortical hierarchy, from acoustic and perceptual features to lexical, compositional and contextual representations (Caucheteux & King, 2022; De Heer et al., 2017), and that model-brain alignment is strongest when models capture hierarchical and contextual computations in higher-order language and association networks (Caucheteux et al., 2023; Mischler et al., 2024; Tang et al., 2023). These findings suggest that early language learning engages multiple representational levels from the outset, but that these levels may differ in their relevance for subsequent learning. Sensory and higher-order linguistic representations were both detectable during Day-1 learning, whereas representations in language and association networks also predicted later learning success. In parallel, model optimization progressively shifted correspondence from sensory toward higher-order components of the same Day-1 neural geometry. Together, these findings suggest that higher-order neural representations expressed during early learning may be particularly informative about subsequent generalization, while progressive model optimization increases correspondence with these same neural components. Whether the relative contribution of these representations changes over the course of human learning remains an open question. This distinction suggests that higher-order linguistic representations need not emerge only at later learning stages to be relevant to successful acquisition; their correspondence with computational models and their prospective behavioral relevance may reflect different aspects of the learning process.

Individualized modeling further showed that early neural information captured by feedback-supervised model representations was informative about individual differences in subsequent generalization. Early behavioral data alone provided only modest predictive power for later generalization, whereas models using early neural data performed significantly better, especially with feedback-related alignment (outcome prediction improved from $r = 0.27$ to $r = 0.43$). This clarifies the relative prediction-feedback dissociation by showing that the signal most strongly aligned with shared neural geometry is not necessarily the one that best predicts individual learning outcomes. Prior work has shown that model-brain alignment varies across individuals, and the degree of alignment can correlate with humans' language comprehension performance (Caucheteux et al., 2022; Schrimpf et al., 2021). Here, we extend that logic to longitudinal learning outcomes. By combining early behavioral responses with participant-specific early neural targets compressed into the model's representational space, we were able to predict later generalization performance more accurately than early behavior alone. This improvement highlights the value of neural information for forecasting future learning outcomes, consistent with previous work showing that early distributed neural patterns contain information about individual language-learning profiles and later achievement (Feng et al., 2021; Morgan-Short, 2020; Song et al., 2026).

Model-brain alignment and behavioral prediction address distinct questions and operate over different temporal scales. Model-brain RSA assesses correspondence between model representations and the neural geometry expressed during Day-1 learning, whereas the behavioral analyses evaluate the prospective relevance of model-derived information for subsequent learning outcomes. GPT-P showed stronger correspondence with the group-level Day-1 neural geometry, suggesting that predictive representations better capture representational structure shared across learners during early acquisition. In contrast, GPT-F-related representations were more informative about later behavioral performance and individual differences in Day-7 generalization. Thus, the computational objective that best captures early shared neural structure need not be the one that is most informative about subsequent behavioral variation. These measures therefore provide complementary constraints on learning rather than interchangeable indices of model efficiency.

More broadly, these findings extend model-brain alignment to language learning by comparing the representational consequences of different computational learning objectives. Previous work has primarily examined correspondence between artificial models and human neural activity during language processing. Here, we combined objective-dependent model comparisons with early-learning neuroimaging and prospective behavioral prediction to provide complementary constraints on computational accounts of how novel linguistic knowledge is acquired. By using a common transformer backbone and the same artificial-language system, we reduced several architectural and stimulus-related differences across models. Nevertheless, the objectives required different supervisory targets and training-input distributions. The observed differences therefore characterize the representational consequences of these training regimes rather than isolating the learning objectives from all other aspects of model experience. This makes it possible to ask not simply which model is more brain-like, but which computational pressure best explains shared neural geometry and learner-specific generalization. Thus, the study extends model-brain alignment from a method for comparing artificial and biological representations to an approach for testing how adults acquire new linguistic structure and why learners exposed to the same input diverge in later outcomes.

Several limitations should be considered when interpreting these findings. First, Brocanto2 offers strong experimental control over input, learners' prior training history, and grammatical structure, but it is still an artificial language with a limited lexicon and a simplified learning context (Morgan-Short et al., 2010, 2012). The extent to which the current findings generalize to naturalistic language learning remains open. Second, our comparison is between matched learning objectives, not between exhaustive implementations of full theoretical frameworks. The findings therefore support complementary roles of prediction-based and feedback-based learning signals as operationalized here, rather than definitively resolving predictive-coding theory or reinforcement learning in general. Third, participants' neural data were collected during early learning, whereas the full behavioral learning experiment unfolded over seven days. Repeated neuroimaging across the full training period would provide a stronger test of how the balance between prediction and feedback signals changes over time. Fourth, although humans and models were matched on materials and overall task structure, they were not matched on all aspects of exposure and learning dynamics, such as the number of repeated exposures required to reach maximum performance. Future work should test whether the same pattern holds with richer linguistic input, longer training windows and more naturalistic tasks. Finally, individual learning outcomes may also depend on prior linguistic and multilingual experience, age, socioeconomic background, motivation, and cognitive abilities, including memory, executive functioning, and reasoning. Our analyses do not disentangle these influences or determine whether particular experiences and cognitive resources compensate for learning difficulties. The present findings therefore identify early neural and behavioral predictors of later generalization without establishing a comprehensive causal account of individual differences. Incorporating these factors in future studies could clarify their contributions to learning outcomes and assess whether the identified predictors generalize across learners with different backgrounds and abilities.

In summary, this study reveals complementary contributions of prediction- and feedback-related signals. Representations optimized for prediction best accounted for the neural geometry shared across learners during early human learning, whereas feedback-related neural signals more strongly predicted variability in later generalization. These findings support a multi-signal framework of adult language learning. More broadly, this work shows that model-brain alignment during learning can move beyond descriptive comparison, providing a principled approach for testing how candidate learning mechanisms shape the acquisition of new structural knowledge and why learners differ in their outcomes.

## Methods

### Participants

One hundred and six healthy adults from the community surrounding South China Normal University (mean age = 21.42 years; 73 females) participated in this study. All were native Mandarin speakers with formal training in English as a second language. None had formally studied a Romance language or been exposed to similar languages for more than three weeks. Four participants started the experiment, but later withdrew. All reported normal hearing and no history of neurological disorders. The study was approved by the Joint Chinese University of Hong Kong (CUHK) - New Territories East Cluster (NTEC) Clinical Research Ethics Committee (CREC) and the Research Ethics Committee at South China Normal University. Informed consent was obtained from each participant. Participants received monetary compensation upon completion.

### Materials

Brocanto2 is a miniature artificial language designed to preserve key properties of natural language while allowing tight experimental control. It is fully productive, can be spoken and comprehended, and embeds lexical and grammatical structures in a meaningful board-game context (Fig. 2a), enabling learners to acquire form-meaning mappings in a constrained yet communicative setting. Previous electrophysiological and neuroimaging studies have used Brocanto (Friederici et al., 2002; Opitz & Friederici, 2003) and Brocanto2 (Morgan-Short et al., 2010, 2012) to investigate the neural correlates of adult artificial-language learning.

The Brocanto2 lexicon comprised 13 lexical items, including four nouns (pleck, neep, blom, vode), two adjectives with gender-marked forms (troise/troiso, neime/neimo), one gender-marked article (li/lu), four verbs (klin, nim, yab, praz) and two adverbs (noyka, zayma). Nouns refer to game tokens and carry formal grammatical roles; verbs refer to actions; adjectives encode token shapes, and adverbs encode movement directions. Articles and adjectives were post-nominal and agreed in gender with the noun, sentences followed a fixed subject-object-verb (SOV) order, and adverbs, when present, immediately followed the verb. To minimize prosodic cues, all words were recorded in isolation and then presented sequentially. In the present study, the materials were organized into three construction types: noun phrases (NPs), simple subject-verb phrases (SVP), and SOV sentences.

### Procedure

Participants completed seven consecutive days of Brocanto2 training. They also completed one auditory and one speech category-learning task for other research purposes. Neuroimaging data were acquired on Day 1, and behavioral training continued across the remaining sessions. Each daily session comprised a vocabulary-exposure phase followed by three grammar-learning tasks targeting NP, SVP, and SOV constructions. On Day 7, participants additionally completed feedback-free generalization tests on novel items. Training and test procedures were implemented in E-Prime (Psychology Software Tools, Inc.) and were adapted from the Brocanto2 learning paradigm used in our earlier work (Feng et al., 2021).

At the start of each session, participants completed a passive vocabulary-exposure phase in which each Brocanto2 lexical item was presented auditorily together with its corresponding visual referent. Grammar learning then proceeded as a grammaticality judgment task with trial-by-trial corrective feedback. No explicit grammatical rules were provided at any point. Instead, participants heard each phrase or sentence while viewing the corresponding game token or movement and judged its acceptability based on their immediate intuition. On-screen feedback then indicated whether the response was correct and served as the learning signal for subsequent trials (Fig. 2a).

Grammar learning trials were organized into six blocks per session, with two blocks for each construction type. Blocks were presented in a fixed order from simpler to more complex structures (i.e., from NP to SVP, then to SOV). Each block included 48 items, half grammatical and half ungrammatical, presented in random order. Ungrammatical items were generated by adding a single violation to an otherwise correct phrase or sentence. Consistent with Brocanto2 morphosyntax, violations targeted gender agreement in NPs and word-order rules in SVP and SOV constructions. During learning, the visual display always matched the intended meaning of the auditory item; thus, the task required grammaticality judgments rather than sentence-picture verification.

Each grammar-learning trial began with a 100-ms fixation cross, followed by a 100-ms blank interval. The auditory stimulus was then presented alongside the corresponding visual display. Stimulus duration varied by construction type: 2,400 ms for NPs, 6,300 ms for SVP items, and 7,600 ms for SOV sentences. After stimulus offset, a question mark prompted the participant to judge grammaticality. Corrective feedback (√ or ×) was displayed for 1,000 ms after the response. To improve the separation of stimulus-related from feedback-related BOLD responses, the interval between response and feedback was jittered from 0 to 4,000 ms, and the inter-trial interval ranged from 2,000 to 6,000 ms.

On Day 7, participants completed a feedback-free grammaticality judgment generalization test assessing transfer of the learned grammar to new Brocanto2 items. In this test, participants judged 144 previously unseen phrases or sentences, including 72 grammatical and 72 ungrammatical items. As in training, each ungrammatical item contained a single violation. The visual displays matched the sentence meaning in all trials, so successful performance depended on identifying grammatical errors rather than semantic mismatches. Participants also completed a picture-sentence mapping test on Day 7 to assess whether spoken Brocanto2 sentences accurately described game tokens or movements. The trial structure matched that of the grammaticality judgment task. Although this task was used to evaluate broader generalization, its data were not included in the current analysis.

## Image acquisition

Neuroimaging data were acquired during the Day-1 language training using a Siemens 3T MRI scanner at the Brain Imaging Center of South China Normal University. Functional and structural images were collected with a 20-channel head coil. Task-based functional images were acquired during the Brocanto2 vocabulary and grammar learning tasks using an interleaved whole-brain echo-planar imaging sequence. Each functional volume contained 56 slices, each 2 mm thick. The acquisition parameters were: repetition time = 2,000 ms, echo time = 30 ms, flip angle = 90°, field of view = $224 \times 224$ $mm^2$, and in-plane resolution = $2 \times 2$ $mm^2$.

High-resolution T1-weighted anatomical images were acquired in the sagittal plane using a magnetization-prepared rapid acquisition gradient-echo sequence, with the following parameters: repetition time = 1,900 ms, echo time = 2.53 ms, flip angle = 9°, and voxel size = $1 \times 1 \times 1$ $mm^3$. Resting-state fMRI and diffusion tensor imaging data were also collected during the same scanning session.

**Brain image preprocessing**

MRI data were preprocessed using fMRIPrep v23.2.1. T1-weighted images were bias-corrected, skull-stripped, segmented into cerebrospinal fluid, white matter, and gray matter, and normalized to MNI space. For each BOLD run, a reference image was coregistered to the T1-weighted image, and head motion was estimated with MCFLIRT. Functional images were resampled to native and MNI space. Nuisance regressors included framewise displacement, DVARS, mean cerebrospinal fluid, white matter, and whole-brain signals, as well as temporal and anatomical CompCor components explaining 50% of the variance. Motion and global signal regressors were expanded with temporal derivatives and quadratic terms, and volumes with framewise displacement greater than 1.0 mm were flagged as motion outliers. All spatial transformations were applied in a single interpolation step using antsApplyTransforms with Lanczos interpolation.

**Language model training**

We trained three customized GPT-style models (i.e., GPT-P, GPT-F and GPT-PF) using Python's *transformers* library v4.42.4. All models shared the same two-layer decoder-only GPT-2 backbone with a hidden size of 768 and a vocabulary of 79 tokens, and differed only in their output heads and training objectives. GPT-P contained 15,024,384 trainable parameters, GPT-F contained 15,025,920 parameters, and GPT-PF contained 15,025,922 parameters. GPT-style models provide a standard autoregressive architecture for next-token prediction, and the *transformers* library also supports sequence-classification heads for GPT-based models. We used byte-pair encoding for tokenization, together with special tokens such as <s> and </s>, yielding a vocabulary of 79 tokens.

GPT-P was trained with a next-token prediction objective using cross-entropy loss. GPT-F was trained with a binary classification head to judge whether each input sequence was grammatical or ungrammatical. GPT-F was also optimized with cross-entropy loss rather than a standard reinforcement-learning objective (e.g., value learning or policy optimization). We therefore refer to it as a feedback-based model because its supervision mirrors the trial-by-trial correctness feedback provided in the human task. In this binary judgment task, explicit correctness feedback served as instructional information (Barto, A. G. & Dietterich, T. G., 2004) that directly specified the desired response rather than functioning only as a scalar reward signal. To distinguish supervised classification from reward-based optimization, we also trained GPT-R using a policy-gradient reinforcement-learning algorithm (Williams, 1992; see Fig. S6).

Finally, the GPT-PF combined the next-token prediction and grammaticality classification objectives within a single model. The combined loss was defined as $L_{combined} = \alpha L_P + L_F$. To prevent the easier binary classification objective from dominating training, the prediction loss was weighted more heavily when generation accuracy lagged behind classification performance. Consistent with this rationale, we used $\alpha = ln(\frac{79}{2})$ when the generation accuracy was lower than the classification accuracy, and $\alpha = 2$ otherwise. This weighting reflects the difference in baseline task difficulty between 79-way next-token prediction and binary grammaticality classification, and also compensates for the smaller amount of grammatical training data available for the prediction objective. Using the same training strategy, we trained a GPT-PR to combine the prediction and reward-based grammaticality classification objectives within a single model.

To assess model training performance, we asked GPT-P to generate outputs for all 1,092 items in the exhaustively constructed Brocanto2 corpus. We evaluated the model's generalization ability as the percentage of generated sentences that were both novel (unseen by the model) and grammatical (i.e., "Gen accuracy"). GPT-F was evaluated using grammaticality-classification accuracy on a held-out 20% validation set, and GPT-PF/PR was evaluated using both metrics.

All three models were trained on datasets intended to approximate the overall scale of human exposure, while differing in the composition of the training input according to their learning objectives. The full training corpus contained grammatical and ungrammatical Brocanto2 sentences in a 1:1 ratio, matching the distribution that human participants encountered during learning. GPT-P was trained only on grammatical sentences, consistent with its generative objective, whereas GPT-F and GPT-PF were trained on 80% of the full corpus, including both grammatical and ungrammatical items. Each model was trained for 200 epochs with a learning rate of $5 \times 10^{-4}$, and the checkpoint with the best validation performance was used for subsequent analyses. Training was conducted on a single NVIDIA RTX A6000 GPU.

As an additional control, we trained a prediction model, GPT-P(L), on a corpus comprising the 144 grammatical Brocanto2 items and minimally corrected versions of the 144 ungrammatical items, thereby matching the number of training items used for GPT-F. All inputs to this control model were therefore grammatical, as in GPT-P, but the corpus size increased from 144 to 288 items. GPT-P(L) showed a training trajectory broadly similar to the main GPT-P model and reached a peak generation accuracy of 0.94 (Supplementary Fig. S5).

**Model performance and changes in internal representations**

To test whether updates in model representations predicted changes in model behavior during training, we analyzed checkpoints prior to the best-performing checkpoint for each model. For each model, we repeated a resampling procedure 1,000 times. In each repetition, 12 checkpoints were randomly sampled. For each checkpoint, all 288 Brocanto2 items were passed through the model, and the final-token hidden-state vector was extracted from each layer, including the embedding layer and the two hidden transformer layers.

For each checkpoint pair, we quantified representational updating by subtracting the item-wise hidden-state vectors at the earlier checkpoint from those at the later checkpoint. This produced a representational-change vector for each item. We then constructed an item-by-item representation-update RDM by computing the Euclidean distance between every pair of representational-change vectors.

Model performance was quantified on the same 288 items and across the same checkpoint pairs. For GPT-P, performance updating was defined as the change in item-level sentence-generation probability, computed as the mean log probability of the sentences. For GPT-F, performance updating was defined as the change in grammaticality-judgment output (i.e., grammatical or ungrammatical). For GPT-PF, performance updating was computed separately for the generation branch and the grammaticality-judgment branch, and also for a combined measure obtained by averaging the z-scored changes from the two branches. For each performance measure, we constructed a performance-update RDM using Euclidean distance.

Finally, for each checkpoint pair and layer, we correlated the representational-update RDM with the corresponding performance-update RDM using Spearman correlation. Within each repetition, correlations were averaged across all checkpoint pairs to obtain one summary value per layer and behavioral measure. The resulting 1,000 resampled values were used to estimate the stability of the representation-performance coupling and to compare this coupling across layers and models.

**Model-brain alignment with representational similarity analysis (RSA)**

We used RSA to compare the geometry of the model and neural representations during language learning. RSA provides a common framework for comparing representational structures across computational models and brain activity, and Euclidean distance is widely used in searchlight-based multivoxel analyses when both spatial patterns and response magnitudes may carry information. For each Brocanto2 phrase or sentence, the full token sequence was passed through each trained model, and the hidden state at the final token was extracted from each model layer (the embedding layer plus two GPT hidden layers) as the sequence representation. We then constructed layer-specific model RDMs across stimuli using cosine distance. To analyze training-related changes in model representations, we selected four checkpoints along each model's performance trajectory, yielding three consecutive training stages. For GPT-P, these checkpoints corresponded to generation accuracies of 0.33, 0.52, 0.71, and 0.91; for GPT-F, they corresponded to GJT accuracies of 0.51, 0.67, 0.83, and 1.00. At each stage, we calculated checkpoint-to-checkpoint difference vectors in hidden-state space and built representational-update RDMs using Euclidean distance, which measures representational change across model training phases.

For the neural data, stimulus-specific *t*-statistic maps were estimated using a least-squares-separate (LSS; Mumford et al., 2012) first-level general linear model implemented in Nilearn (v0.10.3). The LSS analysis included only trials from the Day-1 grammar-learning phase; the vocabulary-exposure phase was not included. A separate GLM was fitted for each target trial. We modeled the target trial with an individual regressor, whereas the remaining grammar-learning trials were modeled using their condition-specific regressors. Event regressors were convolved with the hemodynamic response function, and the model included nuisance regressors and temporal drift terms. A *t*-contrast on the target-trial regressor yielded one trial-specific *t*-statistic map, from which the neural RDMs were constructed.

Whole-brain RSA was implemented using a searchlight procedure in native space, with a 10-mm spherical region centered on each voxel. Within each searchlight, Euclidean distance quantified dissimilarity among local multivoxel activity patterns, and the resulting maps were transformed into standard space. For the between-model comparison, we fitted a multiple-regression RSA model at each searchlight, including the corresponding layer-specific RDMs from GPT-P, GPT-F, GPT-PF, and GPT-R, together with RDMs representing the untrained model, stimulus length, and experimental block. This analysis estimated each model's contribution after accounting for the other model predictors and nuisance variables. Separate analyses examined correspondence between neural RDMs and representations from different layers or successive training stages. In the cross-layer analysis, each neural RDM was regressed onto the two layer-specific RDMs from the same model. In the representational-update analysis, each neural RDM was regressed onto the three stage-specific update RDMs from a single layer. In both analyses, we included three control RDMs: an untrained GPT exposed to the same inputs, to account for representational structure attributable to architecture and stimulus exposure alone; a stimulus-length RDM, to account for variance related to sequence length and associated processing demands; and a block-identity RDM, to account for variance attributable to the experimental block structure. These controls were included to isolate model-brain alignment specific to the learned representational geometry rather than to generic architectural or design-related similarities. Group-level significance was assessed with threshold-free cluster enhancement (TFCE) and permutation testing (10,000 permutations; family-wise error-corrected $p < 0.05$) (Nichols & Holmes, 2002; Smith & Nichols, 2009).

To characterize alignment dynamics across brain networks, we additionally performed network-level RSA as part of the representational update analysis. The brain was parcellated into 12 functional networks using the Ji et al. (2019) atlas. Neural RDMs were computed within each network using Euclidean distance, and alignment with the three stage-specific update RDMs was assessed with partial Pearson correlations. This analysis estimated the unique contribution of each training stage while controlling for the other two stages.

**Robustness across random initializations**

We independently trained 10 instances each of GPT-P, GPT-F, and GPT-R with different random seeds for parameter initialization, while keeping architecture, training materials, and other training settings fixed within each objective. For each hidden layer, representational reproducibility was quantified by correlating the corresponding upper-triangular, off-diagonal RDM entries across all 45 unique pairs of runs. Whole-brain RSA was repeated separately for each run and for the seed-averaged RDMs, retaining the nuisance regressors and statistical inference procedure described above.

Brain-alignment-map reproducibility was quantified using spatial correlations between unthresholded whole-brain RSA maps. We calculated pairwise inter-seed correlations and leave-one-seed-out correlations between each run's map and the mean map of the remaining nine runs. Split-half reliability was assessed by repeatedly dividing the 10 runs into two groups of five, correlating their mean maps, and applying the Spearman-Brown correction. Within-GPT-P map similarity was additionally compared with GPT-P–GPT-F and GPT-P–GPT-R similarity. Confidence intervals for the mean pairwise correlations were estimated by repeating the analysis 10 times, each time leaving out one independently trained model. This accounts for the fact that pairwise correlations sharing the same model are not independent. Cross-objective differences were tested using paired permutation tests that preserved the matching of random seeds across model conditions.

**Individualized neurocomputational modeling with neural, behavioral, and language data**

Unlike the earlier GPT models, which were optimized for corpus-level performance, the individualized models were designed to additionally capture participant-specific behavioral and neural structures to predict later generalization outcomes and to examine whether prediction- or feedback-related neural signals explained individual differences in learning.

To incorporate neural signals into the models, we trained a self-supervised autoencoder to compress each participant's high-dimensional brain data into a lower-dimensional representation matched to the dimensionality of GPT's hidden states. This compressed neural representation was then used as an additional training target. Although linear mapping approaches are commonly used to relate brain activity to computational models, they are limited in capturing nonlinear structure in large-scale neural data. By contrast, the autoencoder-based approach was intended to preserve the underlying nonlinear geometry in neural representations and, crucially, enabled joint optimization of behavioral, neural, and next-token objectives (Han et al., 2019; Huang et al., 2018). ROIs were defined using brain masks derived from significant regions in alignment between late-stage model updates and Day-1 neural representations. Individualized models were then trained using prediction-related brain masks, feedback-related brain masks or their combination, allowing us to test which type of neural information best accounted for inter-learner variability (Fig. 5a).

For each participant, we trained a participant-specific GPT model with three joint objectives: predicting that participant's grammaticality judgments during early learning (Days 1-2), fitting early neural responses from selected brain regions, and predicting the next token in Brocanto2 sequences. Because the neural data in standard space were much higher-dimensional than the model hidden state (271,633 voxels versus 768 units), we first compressed the neural data with the autoencoder (Han et al., 2019; Huang et al., 2018). The autoencoder comprised three fully connected layers, with a 768-dimensional bottleneck matched to the GPT hidden size. For each participant, it was trained on the other 101 participants in a leave-one-subject-out procedure and then applied to the held-out participant to generate low-dimensional neural targets. This approach ensures dimensionality transformation while preserving as much information as possible from the original neural data.

The individualized GPT was optimized with a weighted multitask objective, $L_{\text{combined}} = \alpha L_{\text{behavior}} + \beta L_{\text{neural}} + \gamma L_{\text{predict}}$, where $L_{\text{behavior}}$ was the cross-entropy loss for participant-specific judgment prediction, $L_{\text{neural}}$ was the mean-squared error between the model's final-layer hidden states and the compressed neural data, and $L_{\text{predict}}$ was the cross-entropy loss for next-token prediction. This weighted-sum formulation is standard in multitask learning when objectives differ in type and scale, but the exact coefficients are heuristic balancing hyperparameters rather than quantities with a unique theoretical optimum (Ruder, 2017). We set $\gamma = 1$ as the reference weight for the language-model objective. We set $\alpha = \ln(79/2)$ to upweight the behavioral term relative to next-token prediction, linking the weight to the difference in baseline cross-entropy between a 79-way token-prediction problem and a binary judgment problem. We set $\beta = 4/768$ as a small scaling factor to keep the neural regression term numerically comparable to the two cross-entropy losses. We fixed the coefficients across comparisons without tuning to Day-7 outcomes. The factor 4/768 scaled the neural loss without reducing model capacity. These comparisons evaluated neural constraints under a common weighting scheme, rather than establishing robustness across alternative coefficients. Behavioral supervision used each participant's observed responses on Days 1 and 2; to reduce day-specific overfitting, training alternated between Day 1 and Day 2 response labels every 30 epochs.

For each participant, Day-1 and Day-2 data were split into 70% for training and 30% for held-out validation. Models were trained for 100 epochs, and the checkpoint with the highest held-out behavioral accuracy was retained. We then used the retained individual models to predict grammaticality judgments in the Day-7 generalization task, testing whether early behavioral and neural signals forecast later responses to novel items for each learner. All Day-7 generalization sequences were excluded from the individualized models' language-modeling and behavioral-training datasets, as well as from validation and checkpoint selection. The test therefore evaluated generalization to previously unseen combinations of familiar Brocanto2 lexical items and grammatical structures. Predicted generalization scores were calculated from model responses to these unseen items, so their association with observed outcomes was not directly optimized during training. Given the high model capacity relative to participant-specific data, we evaluated predictive utility through subsequent generalization and comparisons of neural constraints under a common architecture.

**Cross-validated extraction of model-captured neural components**

To reduce dependence on region labels alone, we used nested cross-validated partial least squares regression to estimate neural components captured by the GPT model's hidden states. For each model, stimulus-level hidden-state vectors served as predictors, and participant-specific neural responses served as targets. In each outer fold, we performed feature scaling, PLS fitting, and selection of the number of latent components exclusively on the training data via inner cross-validation. We then applied the fitted mapping to the held-out data to generate out-of-sample, model-captured neural targets. These predicted neural components replaced the raw neural targets in the individualized modeling procedure, while the model architecture, behavioral objective, language-modeling objective, and optimization procedure were held constant across the analyses.

We repeated the procedure using voxel-count-matched masks and a common language-network mask, with the latter providing identical anatomical targets for model representations. Multiple-demand and visual-network responses served as control targets. To adjust for nonverbal intelligence, regression models relating pre-training TONI-4 scores to Day-7 generalization were fitted within the training folds of a five-fold cross-validation and applied to held-out participants. Predictions were evaluated against the resulting held-out residuals. Day-7 outcomes were not used for neural-component extraction or individualized-model training. We additionally evaluated neural constraints associated with late-stage model-update representations; the neural measurements remained restricted to Day 1.

**Simpler predictive baselines**

We used PCA followed by ridge regression or SVR to predict participant-level Day-7 generalization from Day-1 neural signals and Days 1-2 behavioral responses. Analyses were performed separately using GPT-P-aligned and voxel-matched GPT-F-aligned neural features. Outer five-fold cross-validation generated predictions for held-out participants, and inner five-fold cross-validation selected hyperparameters. Preprocessing, PCA, and model fitting were restricted to the corresponding training folds. Performance was quantified by the Spearman correlation between pooled out-of-fold predictions and observed generalization scores.

**Data availability**

De-identified behavioral data, model-derived representations and group-level statistical maps will be deposited on OSF before publication.

**Code availability**

The code for replicating the main analyses of this paper is available at: https://github.com/thedarkkinght/GPTs_learn_Brocanto2

**Author contributions**: Shu.Y. and G.F. designed the research; Shu.Y., X. J., and S.W. performed the research; Shu.Y. and Shao.Y. analyzed the data; Shu.Y. and G.F. wrote the first draft of the paper; Shu.Y., Shao.Y., and G.F. edited the paper; Shu.Y. and G.F. wrote the paper.

# Supplementary Materials

## Supplementary Figures

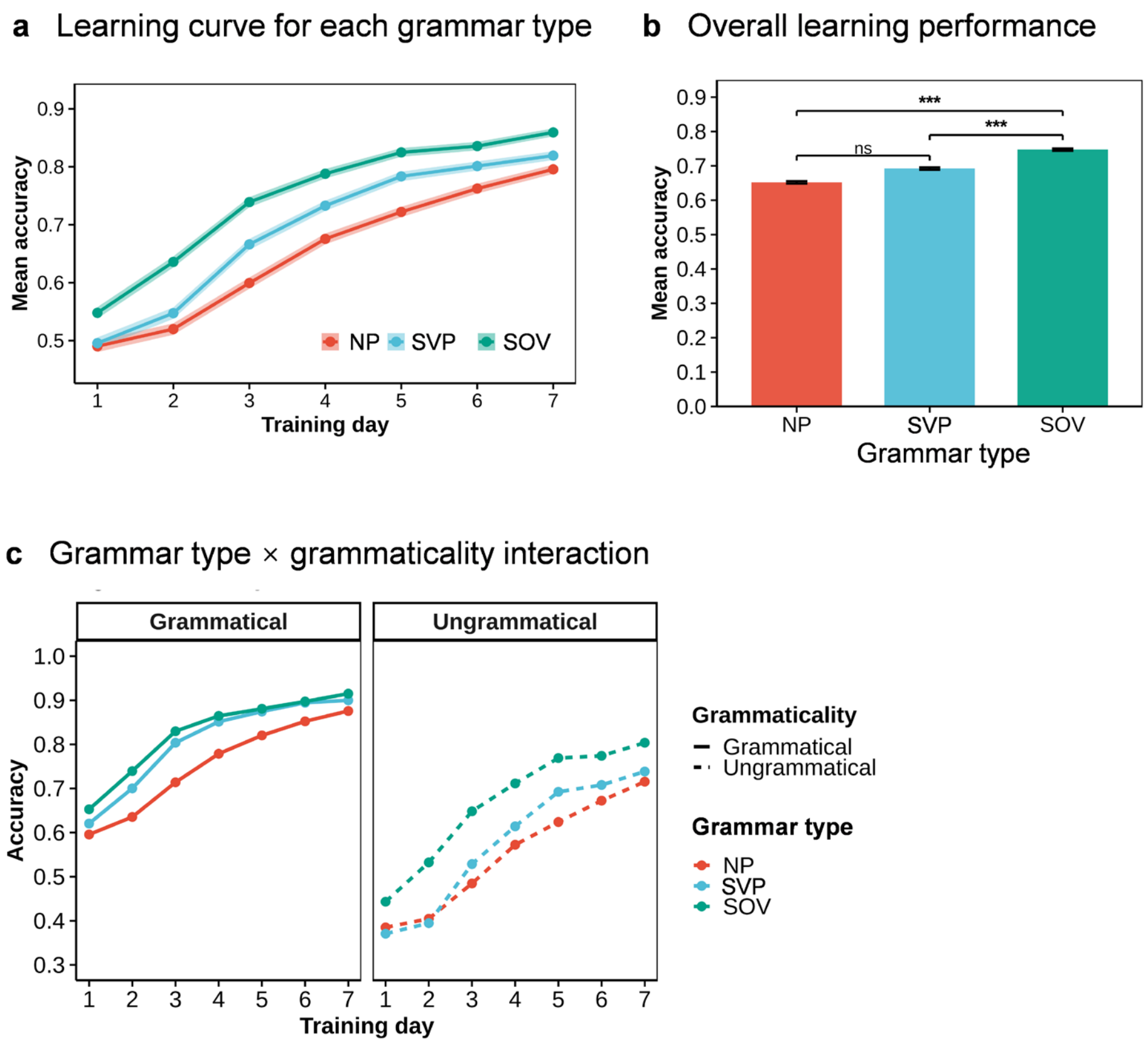

**Fig. S1. Behavioral learning across training days.** Trial-level accuracy was analyzed with a mixed-effects logistic regression including random intercepts for participants and items. **a,** Accuracy improved across the 7-day training period for all construction types, with the highest performance for SOV, followed by SVP and NP. **b,** Main effect of construction type. Pairwise comparisons showed that SOV accuracy was higher than SVP ($\beta = 0.52$, $SE = 0.080$, $z = 6.56$, $p = 5.47 \times 10^{-11}$, $OR = 1.69$) and NP ($\beta = 0.55$, $SE = 0.099$, $z = 5.60$, $p = 1.54 \times 10^{-8}$, $OR = 1.74$), whereas SVP and NP did not differ ($\beta = 0.03$, $SE = 0.099$, $z = 0.31$, $p = 0.76$). **c,** Construction type × grammaticality interaction. Accuracy was higher for grammatical than ungrammatical items overall ($\beta = 1.09$, $SE = 0.13$, $z = 8.34$, $p < 2\times 10^{-16}$, $OR = 2.96$), and this grammaticality effect differed across construction types ($\chi^2_{(2)} = 6.18$, $p = 0.045$). Accuracy also increased reliably across days ($\beta = 0.29$, $SE = 0.006$, $z = 46.74$, $p < 2\times 10^{-16}$), with faster improvement for SVP and SOV than NP (SVP × Day: $\beta = 0.042$, $SE = 0.0088$, z = 4.81, $p = 1.49\times 10^{-6}$; SOV × Day: $\beta = 0.031$, $SE = 0.0090$, $z = 3.48$, $p = 5.01\times 10^{-4}$).

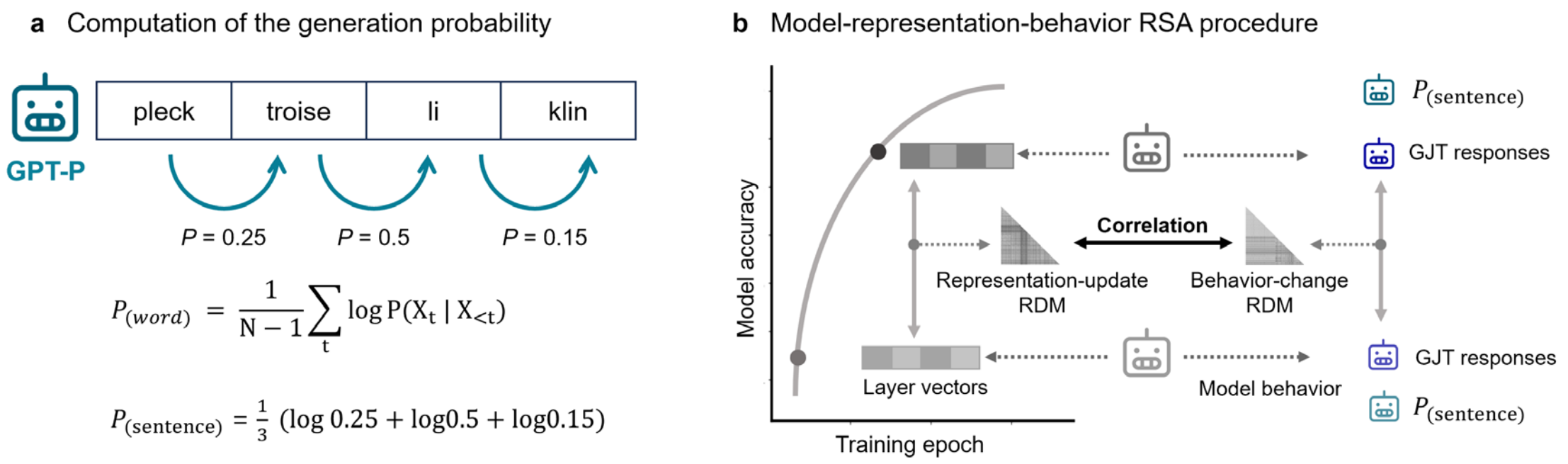

**Fig. S2. Similarity analysis procedure to examine the relationship between model representational updates and behavioral changes. a,** Item-level behavioral scores for GPT-P were computed from the mean sentence log probability, obtained by averaging token-wise conditional log probabilities across each Brocanto2 sequence. This score indexed the model's generative behaviors for each item. **b,** Model-representation-behavior RSA procedure. For each pair of training checkpoints, representational updating was computed as the item-wise difference between layer vectors for each model layer. Behavioral updating was computed using the same checkpoint pair from item-level model outputs: mean sentence log probability for GPT-P, grammaticality judgment responses for GPT-F, and the corresponding generation, classification, and combined scores for GPT-PF. Representational-update and behavior-update RDMs were constructed across the 288 Brocanto2 items using Euclidean distance, and their correspondence was assessed using Spearman's rank correlation.

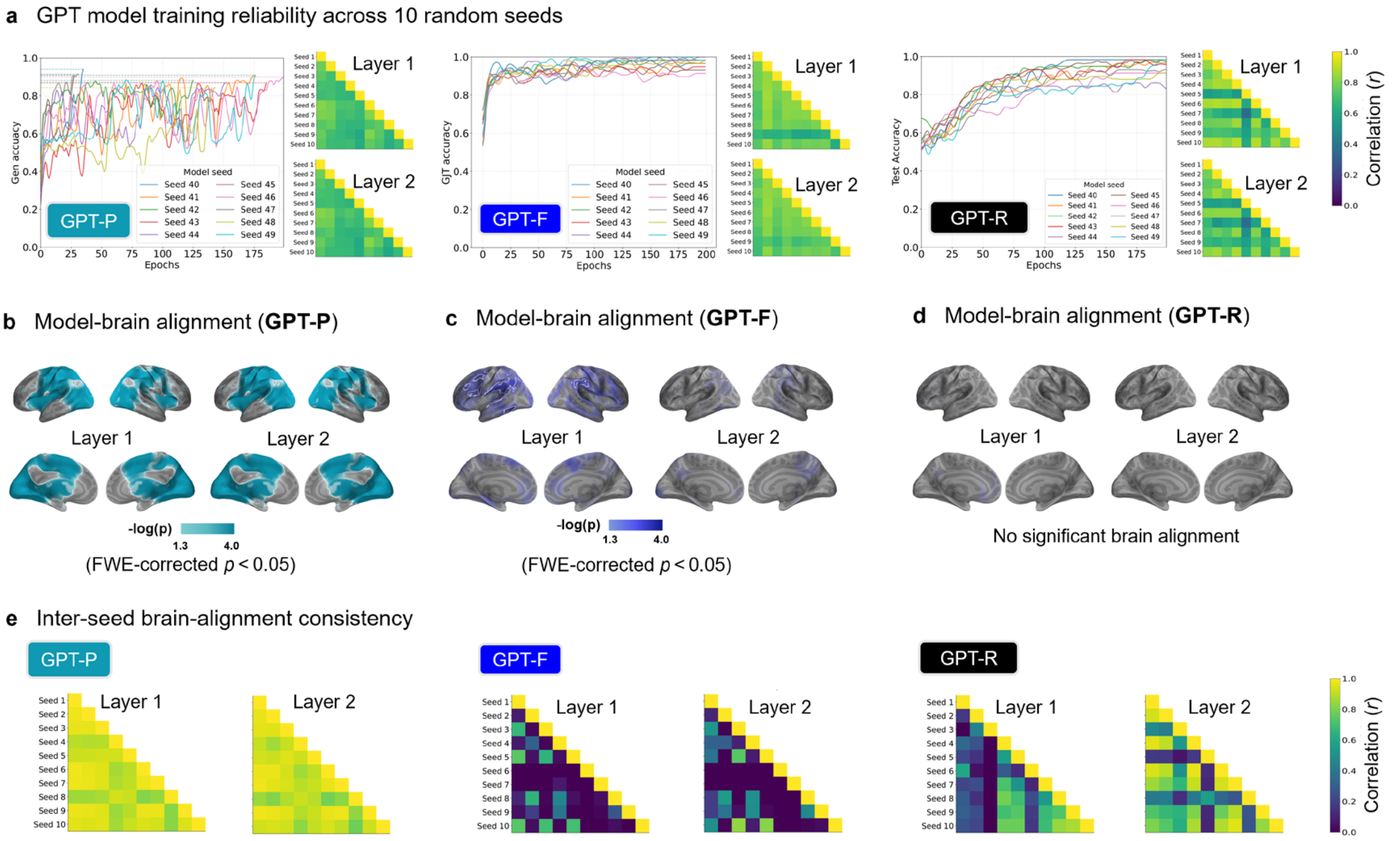

**Fig. S3**. Stability of model representations and model-brain alignment across random seeds. **a**, Training trajectories for GPT-P, GPT-F, and GPT-R, each trained with 10 independent random initializations. Triangular matrices show pairwise inter-seed correlations ($r$) between representational dissimilarity matrices (RDMs) for Layer 1 and Layer 2. **b-d**, Whole-brain representation maps using seed-averaged RDMs for each layer of GPT-P (**b**), GPT-F (**c**), and GPT-R (**d**). GPT-P reproduced broad model-brain alignment across both layers, whereas GPT-F showed a more limited pattern, with significant alignment in Layer 1 but substantially reduced alignment in Layer 2. GPT-R showed no suprathreshold alignment in either layer. **e**, Inter-seed consistency of model-brain alignment. Triangular matrices show pairwise correlations between whole-brain RSA maps across the 10 runs for each model and layer. GPT-P showed highly consistent alignment patterns across seeds in both layers, whereas GPT-F and GPT-R showed greater variability. Separate analyses of the 10 matched GPT-P and GPT-F initializations using the same statistical procedure yielded significant alignment in all 10 GPT-P runs and 4 of 10 GPT-F runs; none of the 10 individual GPT-R runs showed suprathreshold alignment. All statistical brain maps are thresholded at family-wise error (FWE)-corrected $p < 0.05$.

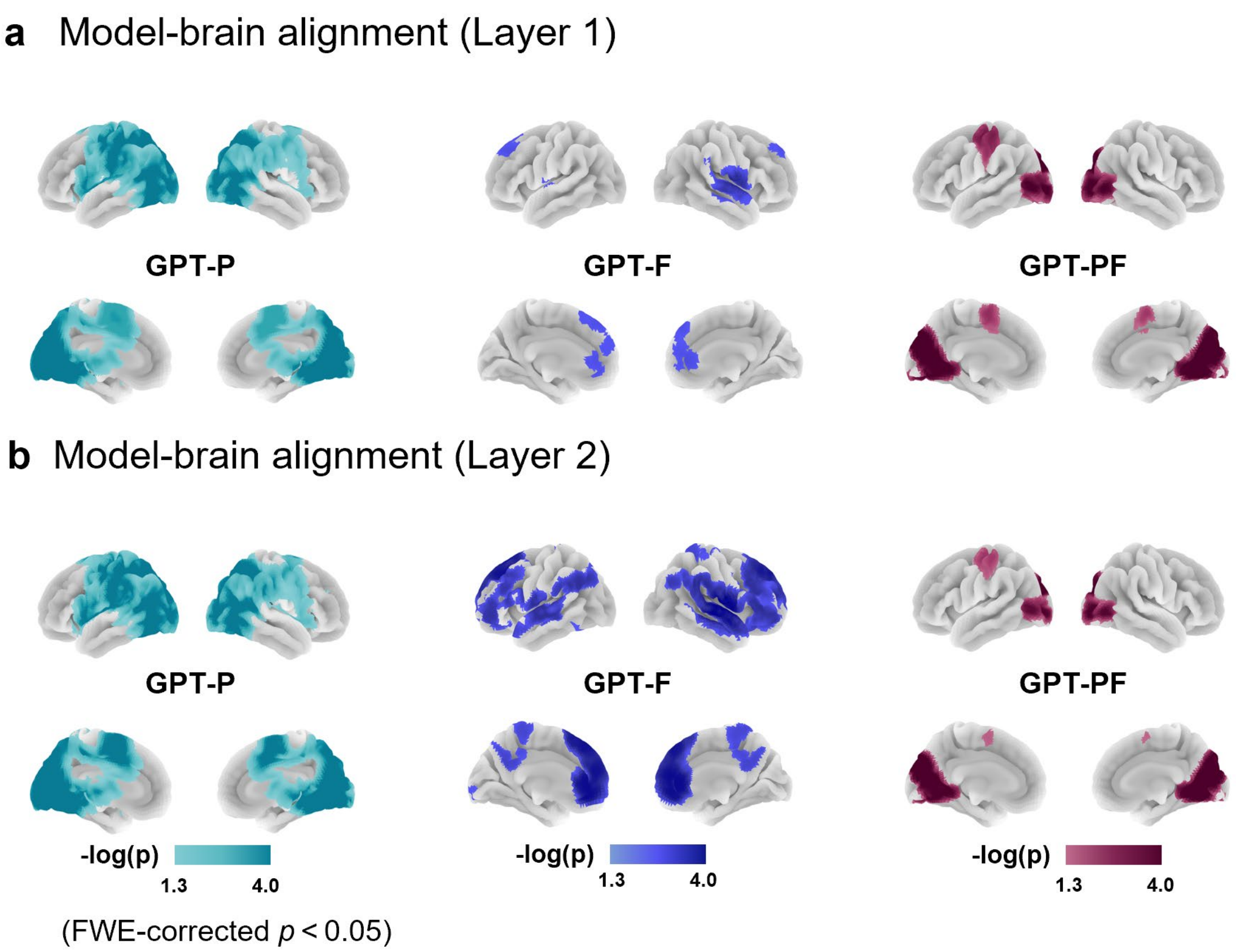

**Fig. S4. Layer-wise model-brain representational alignment. a**,**b**, Whole-brain searchlight RSA results for the first (a) and second (b) hidden layers of GPT-P, GPT-F and GPT-PF. Model RDMs were fitted to fMRI-derived neural RDMs while controlling for an untrained GPT baseline, stimulus length and block identity. GPT-P showed broad alignment across both layers. GPT-F showed weaker, more spatially restricted alignment in layer 1 but stronger alignment in layer 2. GPT-PF showed significant alignment in both layers, with a more focal spatial pattern in sensorimotor regions. Brain maps were thresholded using TFCE with 10,000 permutations and family-wise-error correction at $p < 0.05$. Color bars indicate -log($p$).

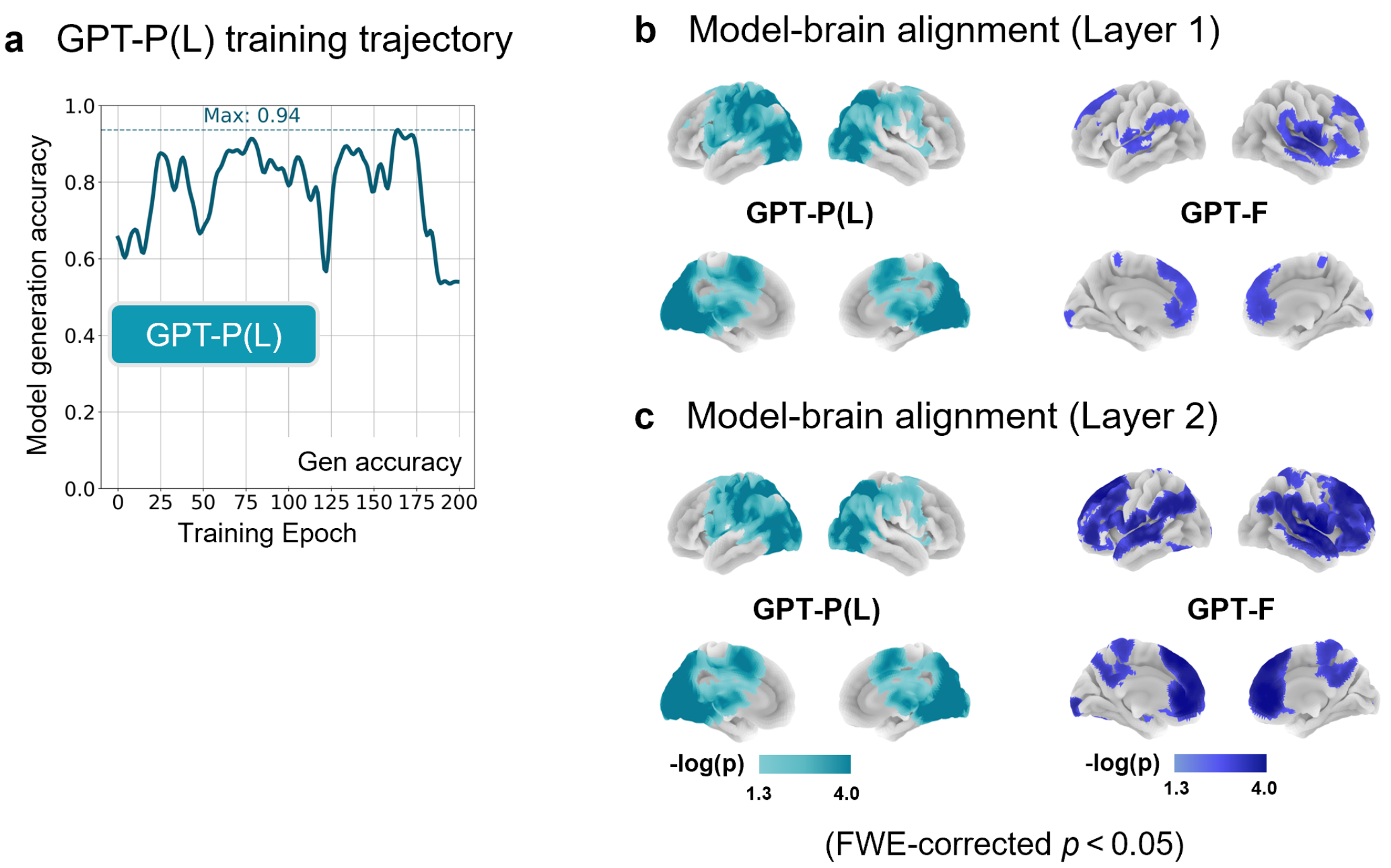

**Fig. S5. Control prediction model trained on an expanded grammatical corpus. a,** training trajectory of GPT-P(L), a prediction model trained on the original 144 grammatical Brocanto2 items plus 144 minimally-corrected grammatical versions of ungrammatical items. GPT-P(L) showed a model training trajectory similar to the main GPT-P model and reached a peak generation accuracy of 0.94. **b, c,** Whole-brain model-brain alignment results for layer 1 (**b**) and layer 2 (**c**). GPT-P(L) was compared with the same GPT-F models used in the main analysis. Consistent with the main prediction-model findings, GPT-P(L) showed robust and spatially extensive alignment with human neural representations across layers. Brain maps were thresholded using TFCE with 10,000 permutations and family-wise-error correction at $p < 0.05$. Color bars indicate -log($p$).

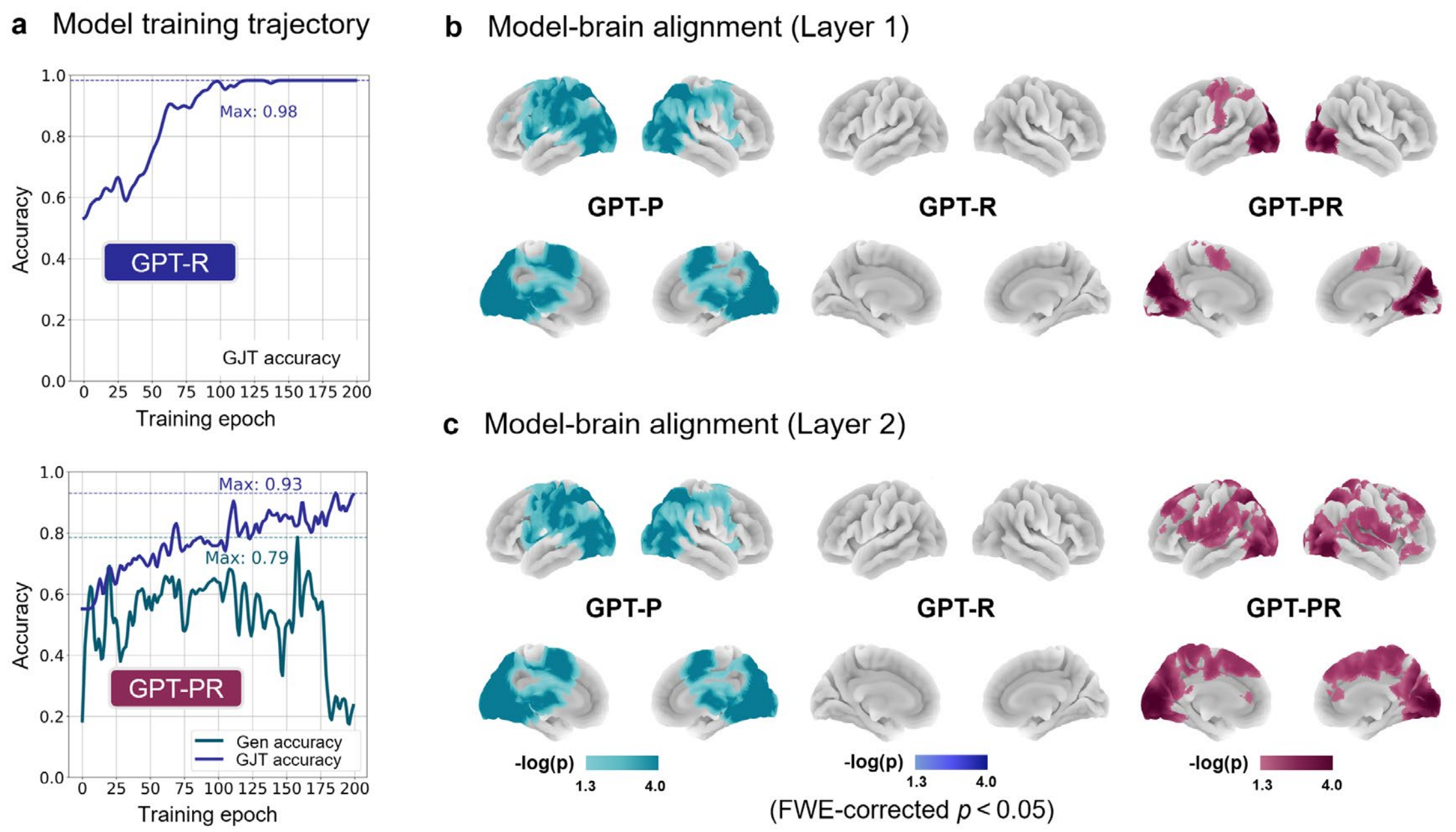

**Fig. S6. Policy-gradient control models and model-brain alignment.** **a**, Validation performance across training epochs for the policy-gradient grammaticality model, GPT-R, and the combined prediction-plus-policy-gradient model, GPT-PR. GPT-R was evaluated by grammaticality-classification accuracy and reached a maximum accuracy of 0.98. GPT-PR was evaluated on both grammaticality classification accuracy and generation accuracy, achieving a maximum classification accuracy of 0.93 and a maximum generation accuracy of 0.79. **b**, **c**, Whole-brain RSA results for the first (**b**) and second (**c**) hidden layers. GPT-P showed significant model-brain alignment in both layers, whereas GPT-R showed no suprathreshold alignment despite high task accuracy. GPT-PR showed significant alignment, particularly in the second layer. Brain maps were thresholded using TFCE with 10,000 permutations and family-wise-error correction at $p < 0.05$. Color bars indicate -log($p$); grey indicates no suprathreshold voxels. These results indicate that policy-gradient task learning alone was insufficient to reproduce human neural representational geometry, whereas alignment emerged when policy-gradient learning was combined with a prediction objective.

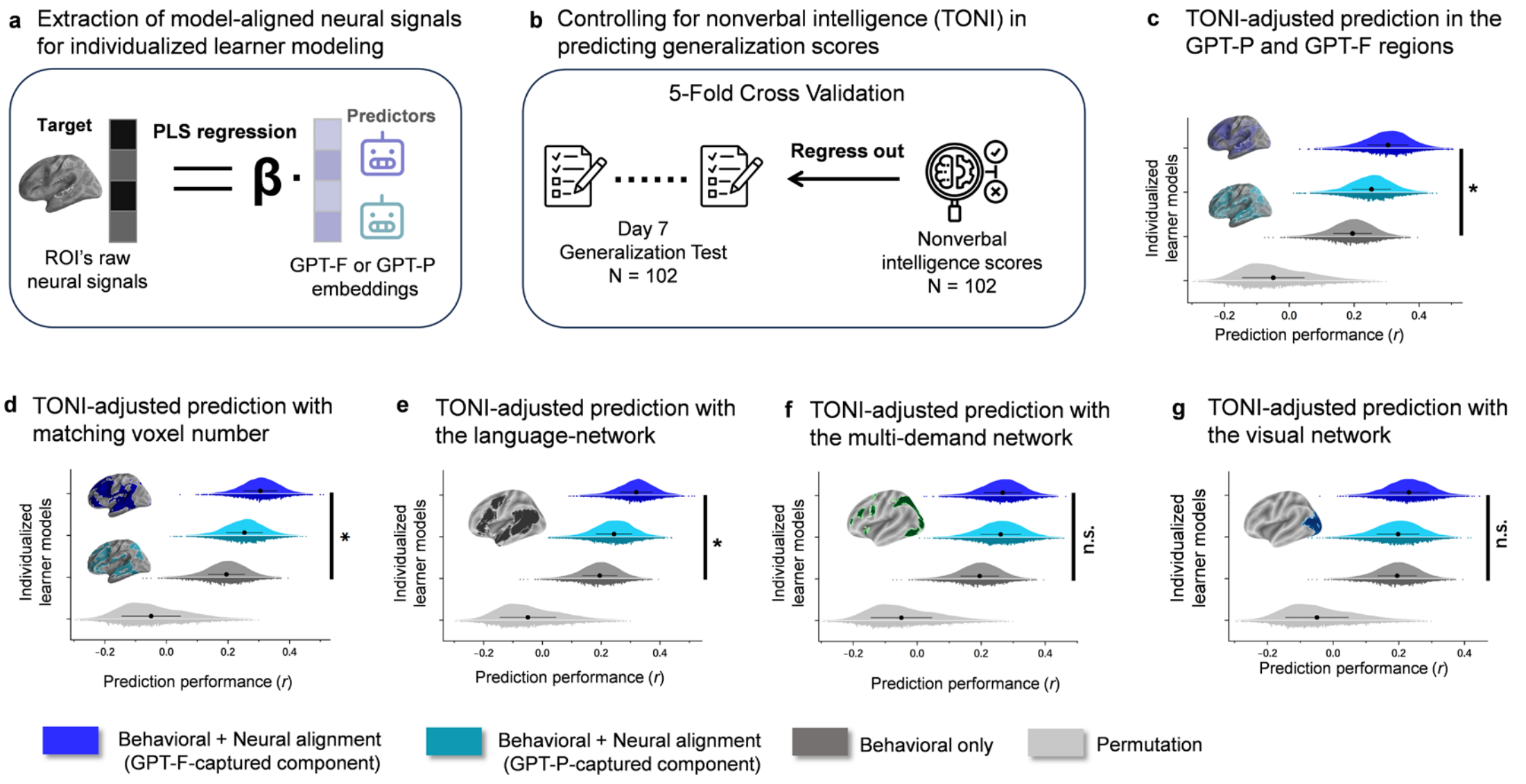

**Fig. S7**. Model-captured neural components differentially contributed to Day-7 generalization prediction after controlling for nonverbal intelligence. **a**, Model-captured neural components were estimated using five-fold cross-validated partial least squares (PLS) regression, with GPT-P or GPT-F embeddings as predictors and trial-wise neural responses as targets. Out-of-sample estimates quantified the neural response components captured by each model and were incorporated into individualized learner models. **b**, Pre-training nonverbal intelligence, measured using the Test of Nonverbal Intelligence, Fourth Edition (TONI-4), was regressed out of Day-7 generalization scores using five-fold cross-validation. **c**, Prediction of TONI-adjusted generalization using neural components derived from GPT-P- and GPT-F-aligned regions. GPT-F-informed model predictions correlated significantly with adjusted generalization scores (Spearman $r = 0.34$, $p = 4.57 \times 10^{-4}$). **d**,**e**, Prediction remained significant after matching voxel counts between GPT-aligned regions (**d**; $r = 0.35$, $p = 3.34 \times 10^{-4}$) and when both models' neural components were derived within a common language-network mask (Wolna et al., 2025), holding anatomical space constant (**e**; $r = 0.36$, $p = 1.90 \times 10^{-4}$). **f**,**g**, Control analyses used neural components derived from the multiple-demand network (Lipkin et al., 2022) and visual network (Schaefer et al., 2018). Although model predictions correlated significantly with adjusted generalization scores (multiple-demand: $r = 0.29$, $p = 0.0036$; visual: $r = 0.23$, $p = 0.0184$), neither control network significantly improved prediction over the behavioral-only model. Raincloud plots show distributions of prediction performance for individualized models incorporating behavioral responses and GPT-F-captured neural components (blue), behavioral responses and GPT-P-captured neural components (teal), behavioral alone (dark gray), and a permutation baseline (light gray). Asterisks denote significant between-model comparisons: * $p < 0.05$; n.s., not significant.

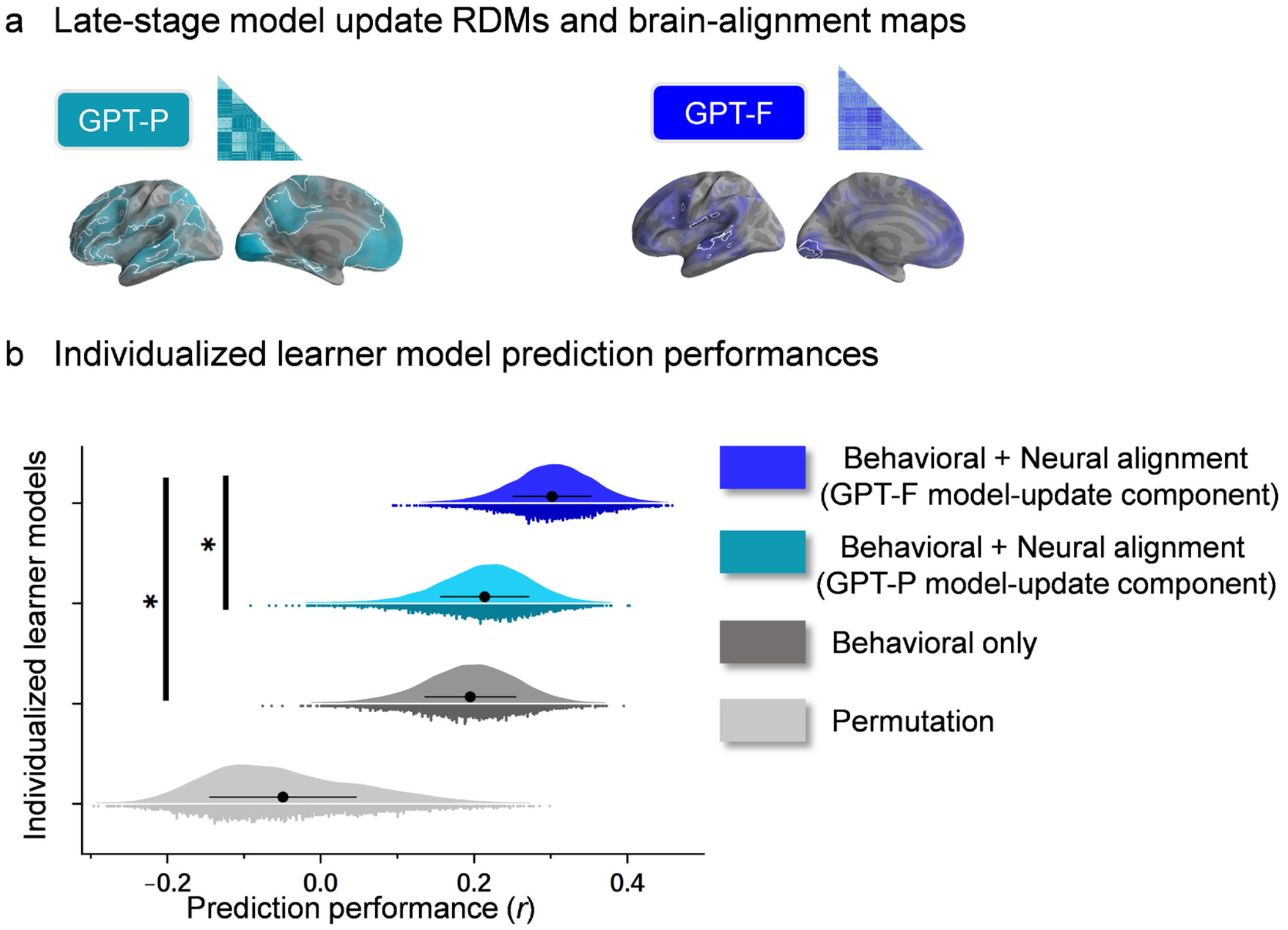

**Fig. S8**. Neural alignment with late-stage model-update representations improves prediction of Day-7 generalization. **a**, Representational dissimilarity matrices (RDMs) of late-stage model updates for GPT-P (teal) and GPT-F (blue), together with their corresponding brain-alignment maps. Late stage refers to model training. **b**, Prediction performance for individualized learner models incorporating behavioral alignment and neural constraints derived from GPT-F model-update components (blue), GPT-P model-update components (teal), alongside a behavioral-only model (dark gray) and a permutation baseline (light gray). Raincloud plots show distributions of correlations between model-predicted and observed Day-7 generalization scores. Incorporating GPT-F-related neural constraints yielded the strongest prediction and significantly improved performance over behavioral alignment alone and GPT-P-related neural constraints. Asterisks indicate significant between-model comparisons (*$p < 0.05$).

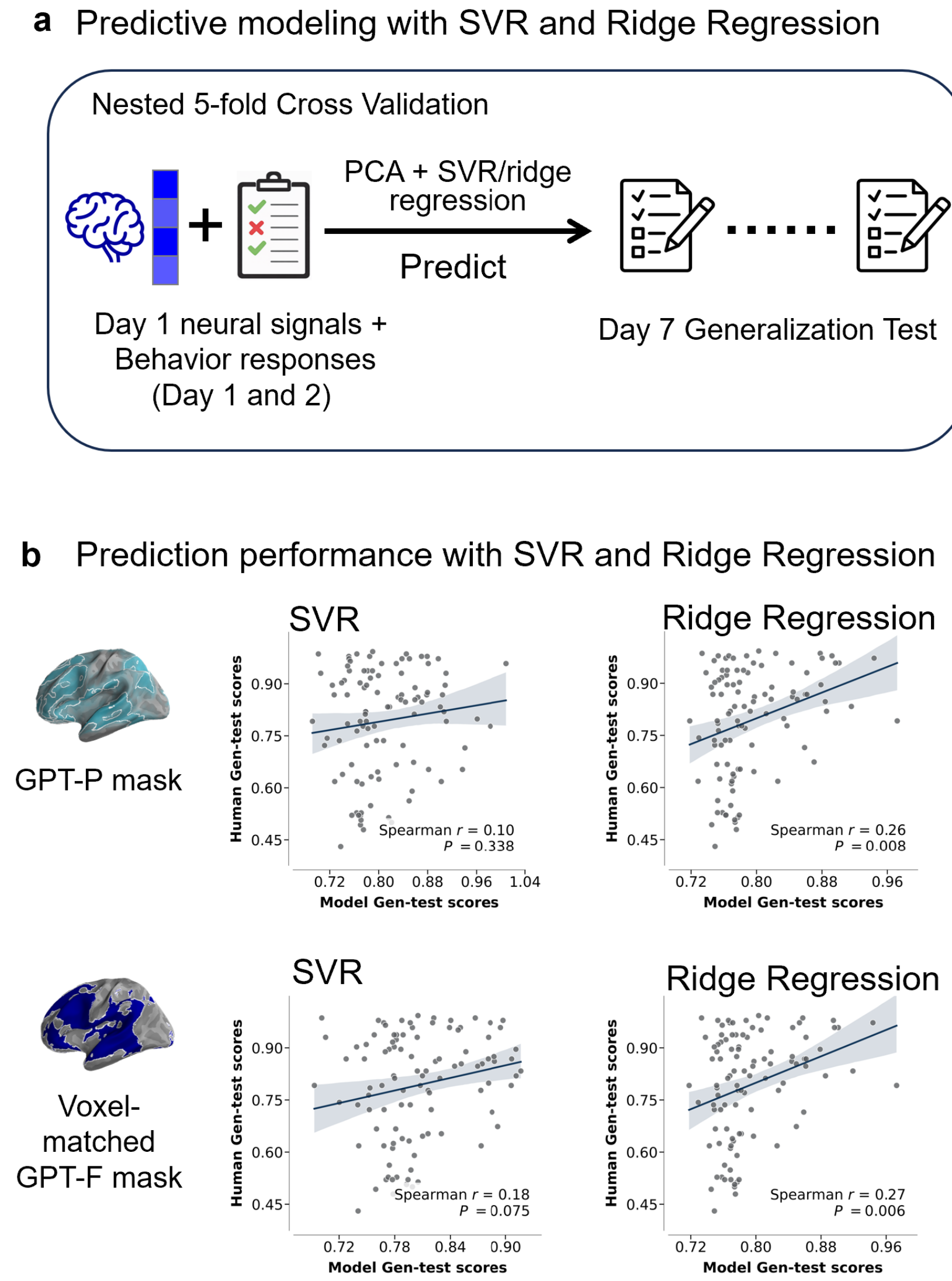

**Fig. S9**. Ridge-regression and support vector regression baselines for predicting Day-7 generalization. a, Schematic of the nested five-fold cross-validation procedure. Day-1 neural signals and behavioral responses from Days 1 and 2 served as predictors of Day-7 generalization. Principal component analysis (PCA), hyperparameter selection, and model fitting were restricted to the training folds, with prediction evaluated on held-out participants. b, Associations between model-predicted and observed generalization scores using neural features from the GPT-P-aligned mask (top) and voxel-matched GPT-F-aligned mask (bottom). Ridge regression yielded modest but significant predictions for both feature spaces (GPT-P: Spearman $r$ = 0.26, $p$ = 0.008; voxel-matched GPT-F: $r$ = 0.27, $p$ = 0.006). Support vector regression (SVR) did not yield significant predictions (GPT-P: $r$ = 0.10, $p$ = 0.338; voxel-matched GPT-F: $r$ = 0.18, $p$ = 0.075). Each point represents one participant; lines indicate fitted linear trends.

## References

Arbel, Y., Hong, L., Baker, T. E., & Holroyd, C. B. (2017). It's all about timing: An electrophysiological examination of feedback-based learning with immediate and delayed feedback. *Neuropsychologia*, *99*, 179–186. https://doi.org/10.1016/j.neuropsychologia.2017.03.003

Arbel, Y., Murphy, A., & Donchin, E. (2014). On the utility of positive and negative feedback in a paired-associate learning task. *Journal of Cognitive Neuroscience*, *26*(7), 1445–1453. https://doi.org/10.1162/jocn_a_00617

Ashby, F. G., Alfonso-Reese, L. A., Turken, A. U., & Waldron, E. M. (1998). A neuropsychological theory of multiple systems in category learning. *Psychological Review*, *105*(3), 442–481. https://doi.org/10.1037/0033-295x.105.3.442

Barto, A. G. & Dietterich, T. G. (2004). Reinforcement learning and its relationship to supervised learning. *Handbook of Learning and Approximate Dynamic Programming*. https://doi.org/10.1109/9780470544785.ch2

Bovolenta, G., & Marsden, E. (2022). Prediction and error-based learning in L2 processing and acquisition: A conceptual review. *Studies in Second Language Acquisition*, *44*(5), 1384–1409. https://doi.org/10.1017/S0272263121000723

Brincat, S. L., Siegel, M., Von Nicolai, C., & Miller, E. K. (2018). Gradual progression from sensory to task-related processing in cerebral cortex. *Proceedings of the National Academy of Sciences*, *115*(30), 1–10. https://doi.org/10.1073/pnas.1717075115

Caucheteux, C., Gramfort, A., & King, J.-R. (2022). Deep language algorithms predict semantic comprehension from brain activity. *Scientific Reports*, *12*(1), 16327. https://doi.org/10.1038/s41598-022-20460-9

Caucheteux, C., Gramfort, A., & King, J.-R. (2023). Evidence of a predictive coding hierarchy in the human brain listening to speech. *Nature Human Behaviour*, *7*(3), 430–441. https://doi.org/10.1038/s41562-022-01516-2

Caucheteux, C., & King, J.-R. (2022). Brains and algorithms partially converge in natural language processing. *Communications Biology*, *5*(1), 134. https://doi.org/10.1038/s42003-022-03036-1

Chang, F., Dell, G. S., & Bock, K. (2006). Becoming syntactic. *Psychological Review*, *113*(2), 234–272. https://doi.org/10.1037/0033-295X.113.2.234

Chang, F., Kidd, E., & Rowland, C. F. (2013). Prediction in processing is a by-product of language learning. *Behavioral and Brain Sciences*, *36*(4), 350–351. Cambridge Core. https://doi.org/10.1017/S0140525X12002518

Courellis, H. S., Minxha, J., Cardenas, A. R., Kimmel, D. L., Reed, C. M., Valiante, T. A., Salzman, C. D., Mamelak, A. N., Fusi, S., & Rutishauser, U. (2024). Abstract representations emerge in human hippocampal neurons during inference. *Nature*, *632*(8026), 841–849. https://doi.org/10.1038/s41586-024-07799-x

De Heer, W. A., Huth, A. G., Griffiths, T. L., Gallant, J. L., & Theunissen, F. E. (2017). The hierarchical cortical organization of human speech processing. *The Journal of Neuroscience*, *37*(27), 6539–6557. https://doi.org/10.1523/JNEUROSCI.3267-16.2017

Elman, J. L. (1990). Finding Structure in Time. *Cognitive Science*, *14*(2), 179–211. https://doi.org/10.1207/s15516709cog1402_1

Feng, G., Ou, J., Gan, Z., Jia, X., Meng, D., Wang, S., & Wong, P. C. M. (2021). Neural

fingerprints underlying individual language learning profiles. *The Journal of Neuroscience*, *41*(35), 7372–7387. https://doi.org/10.1523/JNEUROSCI.0415-21.2021

Friederici, A. D., Steinhauer, K., & Pfeifer, E. (2002). Brain signatures of artificial language processing: Evidence challenging the critical period hypothesis. *Proceedings of the National Academy of Sciences*, *99*(1), 529–534. https://doi.org/10.1073/pnas.012611199

Friston, K. (2005). A theory of cortical responses. *Philosophical Transactions of the Royal Society B: Biological Sciences*, *360*(1456), 815–836. https://doi.org/10.1098/rstb.2005.1622

Goldstein, A., Zada, Z., Buchnik, E., Schain, M., Price, A., Aubrey, B., Nastase, S. A., Feder, A., Emanuel, D., Cohen, A., Jansen, A., Gazula, H., Choe, G., Rao, A., Kim, C., Casto, C., Fanda, L., Doyle, W., Friedman, D., … Hasson, U. (2022). Shared computational principles for language processing in humans and deep language models. *Nature Neuroscience*, *25*(3), 369–380. https://doi.org/10.1038/s41593-022-01026-4

Groen, I. I., Greene, M. R., Baldassano, C., Fei-Fei, L., Beck, D. M., & Baker, C. I. (2018). Distinct contributions of functional and deep neural network features to representational similarity of scenes in human brain and behavior. *eLife*, *7*, e32962. https://doi.org/10.7554/eLife.32962

Hagoort, P. (2019). The neurobiology of language beyond single-word processing. *Science*, *366*(6461), 55–58. https://doi.org/10.1126/science.aax0289

Hamrick, P., Lum, J. A. G., & Ullman, M. T. (2018). Child first language and adult second language are both tied to general-purpose learning systems. *Proceedings of the National Academy of Sciences*, *115*(7), 1487–1492. https://doi.org/10.1073/pnas.1713975115

Han, K., Wen, H., Shi, J., Lu, K.-H., Zhang, Y., Fu, D., & Liu, Z. (2019). Variational autoencoder: An unsupervised model for encoding and decoding fMRI activity in visual cortex. *NeuroImage*, *198*, 125–136. https://doi.org/10.1016/j.neuroimage.2019.05.039

Heilbron, M., Armeni, K., Schoffelen, J.-M., Hagoort, P., & De Lange, F. P. (2022). A hierarchy of linguistic predictions during natural language comprehension. *Proceedings of the National Academy of Sciences*, *119*(32), e2201968119. https://doi.org/10.1073/pnas.2201968119

Huang, H., Hu, X., Zhao, Y., Makkie, M., Dong, Q., Zhao, S., Guo, L., & Liu, T. (2018). Modeling Task fMRI Data Via Deep Convolutional Autoencoder. *IEEE Transactions on Medical Imaging*, *37*(7), 1551–1561. https://doi.org/10.1109/TMI.2017.2715285

Ito, A., Pickering, M. J., & Corley, M. (2018). Investigating the time-course of phonological prediction in native and non-native speakers of English: A visual world eye-tracking study. *Journal of Memory and Language*, *98*, 1–11. https://doi.org/10.1016/j.jml.2017.09.002

Ji, J. L., Spronk, M., Kulkarni, K., Repovš, G., Anticevic, A., & Cole, M. W. (2019). Mapping the human brain's cortical-subcortical functional network organization. *NeuroImage*, *185*, 35–57. https://doi.org/10.1016/j.neuroimage.2018.10.006

Johnston, W. J., & Fusi, S. (2023). Abstract representations emerge naturally in neural networks trained to perform multiple tasks. *Nature Communications*, *14*(1), 1–18. https://doi.org/10.1038/s41467-023-36583-0

Kidd, E., Donnelly, S., & Christiansen, M. H. (2018). Individual Differences in Language Acquisition and Processing. *Trends in Cognitive Sciences*, *22*(2), 154–169. https://doi.org/10.1016/j.tics.2017.11.006

Kriegeskorte, N. (2008). Representational similarity analysis – connecting the branches of systems neuroscience. *Frontiers in Systems Neuroscience*. https://doi.org/10.3389/neuro.06.004.2008

Kriegeskorte, N., & Kievit, R. A. (2013). Representational geometry: Integrating cognition, computation, and the brain. *Trends in Cognitive Sciences*, *17*(8), 401–412. https://doi.org/10.1016/j.tics.2013.06.007

Lillicrap, T. P., Santoro, A., Marris, L., Akerman, C. J., & Hinton, G. (2020). Backpropagation and the brain. *Nature Reviews Neuroscience*, *21*(6), 335–346. https://doi.org/10.1038/s41583-020-0277-3

Lipkin, B., Tuckute, G., Affourtit, J., Small, H., Mineroff, Z., Kean, H., Jouravlev, O., Rakocevic, L., Pritchett, B., Siegelman, M., Hoeflin, C., Pongos, A., Blank, I. A., Struhl, M. K., Ivanova, A., Shannon, S., Sathe, A., Hoffmann, M., Nieto-Castañón, A., & Fedorenko, E. (2022). Probabilistic atlas for the language network based on precision fMRI data from >800 individuals. *Scientific Data*, *9*(1), 529. https://doi.org/10.1038/s41597-022-01645-3

Menon, V. (2023). 20 years of the default mode network: A review and synthesis. *Neuron*, *111*(16), 2469–2487. https://doi.org/10.1016/j.neuron.2023.04.023

Minda, J. P., Roark, C. L., Kalra, P., & Cruz, A. (2024). Single and multiple systems in categorization and category learning. *Nature Reviews Psychology*, *3*(8), 536–551. https://doi.org/10.1038/s44159-024-00336-7

Mischler, G., Li, Y. A., Bickel, S., Mehta, A. D., & Mesgarani, N. (2024). Contextual feature extraction hierarchies converge in large language models and the brain. *Nature Machine Intelligence*, *6*(12), 1467–1477. https://doi.org/10.1038/s42256-024-00925-4

Morgan-Short, K. (2020). Insights into the neural mechanisms of becoming bilingual: A brief synthesis of second language research with artificial linguistic systems. *Bilingualism: Language and Cognition*, *23*(1), 87–91. https://doi.org/10.1017/S1366728919000701

Morgan-Short, K., Sanz, C., Steinhauer, K., & Ullman, M. T. (2010). Second Language Acquisition of Gender Agreement in Explicit and Implicit Training Conditions: An Event-Related Potential Study. *Language Learning*, *60*(1), 154–193. https://doi.org/10.1111/j.1467-9922.2009.00554.x

Morgan-Short, K., Steinhauer, K., Sanz, C., & Ullman, M. T. (2012). Explicit and Implicit Second Language Training Differentially Affect the Achievement of Native-like Brain Activation Patterns. *Journal of Cognitive Neuroscience*, *24*(4), 933–947. https://doi.org/10.1162/jocn_a_00119

Mumford, J. A., Turner, B. O., Ashby, F. G., & Poldrack, R. A. (2012). Deconvolving BOLD activation in event-related designs for multivoxel pattern classification analyses. *NeuroImage*, *59*(3), 2636–2643. https://doi.org/10.1016/j.neuroimage.2011.08.076

Naselaris, T., Kay, K. N., Nishimoto, S., & Gallant, J. L. (2011). Encoding and decoding in fMRI. *NeuroImage*, *56*(2), 400–410. https://doi.org/10.1016/j.neuroimage.2010.07.073

Nichols, T. E., & Holmes, A. P. (2002). Nonparametric permutation tests for functional neuroimaging: A primer with examples. *Human Brain Mapping*, *15*(1), 1–25. https://doi.org/10.1002/hbm.1058

Opitz, B., Ferdinand, N. K., & Mecklinger, A. (2011). Timing matters: The impact of immediate and delayed feedback on artificial language learning. *Frontiers in Human Neuroscience*,

*5*. https://doi.org/10.3389/fnhum.2011.00008

Opitz, B., & Friederici, A. D. (2003). Interactions of the hippocampal system and the prefrontal cortex in learning language-like rules. *NeuroImage*, *19*(4), 1731–1737. https://doi.org/10.1016/S1053-8119(03)00170-8

Pelucchi, B., Hay, J. F., & Saffran, J. R. (2009). Learning in reverse: Eight-month-old infants track backward transitional probabilities. *Cognition*, *113*(2), 244–247. https://doi.org/10.1016/j.cognition.2009.07.011

Rabagliati, H., Gambi, C., & Pickering, M. J. (2015). Learning to predict or predicting to learn? *Language, Cognition and Neuroscience*, *31*(1), 94–105. https://doi.org/10.1080/23273798.2015.1077979

Riesenhuber, M., & Poggio, T. (1999). Hierarchical models of object recognition in cortex. *Nature Neuroscience*, *2*(11), 1019–1025. https://doi.org/10.1038/14819

Ruder, S. (2017). *An Overview of Multi-Task Learning in Deep Neural Networks* (arXiv:1706.05098). arXiv. https://doi.org/10.48550/arXiv.1706.05098

Saffran, J. R. (2002). Constraints on statistical language learning. *Journal of Memory and Language*, *47*(1), 172–196. https://doi.org/10.1006/jmla.2001.2839

Saffran, J. R., Aslin, R. N., & Newport, E. L. (1996). Statistical Learning by 8-Month-Old Infants. *Science*, *274*(5294), 1926–1928. https://doi.org/10.1126/science.274.5294.1926

Schaefer, A., Kong, R., Gordon, E. M., Laumann, T. O., Zuo, X.-N., Holmes, A. J., Eickhoff, S. B., & Yeo, B. T. T. (2018). Local-Global Parcellation of the Human Cerebral Cortex from Intrinsic Functional Connectivity MRI. *Cerebral Cortex*, *28*(9), 3095–3114. https://doi.org/10.1093/cercor/bhx179

Schrimpf, M., Blank, I. A., Tuckute, G., Kauf, C., Hosseini, E. A., Kanwisher, N., Tenenbaum, J. B., & Fedorenko, E. (2021). The neural architecture of language: Integrative modeling converges on predictive processing. *Proceedings of the National Academy of Sciences*, *118*(45), e2105646118. https://doi.org/10.1073/pnas.2105646118

Smith, S., & Nichols, T. (2009). Threshold-free cluster enhancement: Addressing problems of smoothing, threshold dependence and localisation in cluster inference. *NeuroImage*, *44*(1), 83–98. https://doi.org/10.1016/j.neuroimage.2008.03.061

Song, P., Yang, S., Geng, X., Gan, Z., Wang, S., & Feng, G. (2026). Multi-network topology underlying individual language learning success. *The Journal of Neuroscience*, *46*(27), e2205252026. https://doi.org/10.1523/JNEUROSCI.2205-25.2026

Stolier, R. M., & Freeman, J. B. (2016). Neural pattern similarity reveals the inherent intersection of social categories. *Nature Neuroscience*, *19*(6), 795–797. https://doi.org/10.1038/nn.4296

Sutton, R. S., & Barto, A. G. (2018). *Reinforcement learning: An introduction*. The MIT Press.

Tang, J., LeBel, A., Jain, S., & Huth, A. G. (2023). Semantic reconstruction of continuous language from non-invasive brain recordings. *Nature Neuroscience*, *26*(5), 858–866. https://doi.org/10.1038/s41593-023-01304-9

Ullman, M. T. (2004). Contributions of memory circuits to language: The declarative/procedural model. *Cognition*, *92*(1–2), 231–270. https://doi.org/10.1016/j.cognition.2003.10.008

Williams, R. J. (1992). Simple statistical gradient-following algorithms for connectionist reinforcement learning. *Machine Learning*, *8*(3), 229–256.

https://doi.org/10.1007/BF00992696

Wolna, A., Wright, A., Casto, C., Lipkin, B., & Fedorenko, E. (2025). The extended language network: Language selective brain areas whose contributions to language remain to be discovered. *bioRxiv*, *2025–04*. https://doi.org/10.1101/2025.04.02.646835